\def\draftversion{false}

\RequirePackage{ifthen}
\ifthenelse{\equal{\draftversion}{true}}{
  \documentclass[aps,prl,galley,showpacs,preprintnumbers,citeautoscript,
      amsmath,amssymb,longbibliography]{revtex4-1}
}{
  \documentclass[citeautoscript,floatfix,aps,prl,twocolumn,
      superscriptaddress,longbibliography]{revtex4-1}
}

\usepackage{amsmath}
\usepackage{graphicx}
\usepackage{epstopdf}
\usepackage{natbib}
\usepackage{array}
\usepackage{dcolumn}
\usepackage{bm}
\usepackage{multirow}
\usepackage{soul}  

\usepackage[usenames,dvipsnames]{color}

\newcommand\T{\rule{0pt}{2.6ex}}              
\newcommand\B{\rule[-1.2ex]{0pt}{0pt}}        

\soulregister\cite7

\ifthenelse{\equal{\draftversion}{true}}{
  \usepackage{showlabels}
  \marginparwidth 2.7in
  \marginparsep 0.5in
  \newcounter{comm} 
  \def\commnext{\stepcounter{comm}}
  \def\commtext{{\bf\color{blue}[\arabic{comm}]}}
  \def\commmar{{\bf\color{blue}[\arabic{comm}]}}
  \def\dvm#1{\commnext\marginpar{\small DV\commmar: #1}\commtext}
  \def\cdm#1{\commnext\marginpar{\small CED\commmar: #1}\commtext}
  \def\msm#1{\commnext\marginpar{\small MS\commmar: #1}\commtext}
  \def\asm#1{\commnext\marginpar{\small AS\commmar: #1}\commtext}
  \def\miq#1{\commnext\marginpar{\small MR\commmar: #1}\commtext}
  \def\mlab#1{\marginpar{\small\bf #1}}
  
}{
  \def\dvm#1{}
  \def\cdm#1{}
  \def\msm#1{}
  \def\asm#1{}
  \def\miq#1{}
  \def\mlab#1{}
  
}

\begin{document}

\title{Direct and converse flexoelectricity in two-dimensional materials}

\author{Matteo Springolo}
\affiliation{Institut de Ci\`encia de Materials de Barcelona 
(ICMAB-CSIC), Campus UAB, 08193 Bellaterra, Spain}

\author{Miquel Royo}
\affiliation{Institut de Ci\`encia de Materials de Barcelona 
(ICMAB-CSIC), Campus UAB, 08193 Bellaterra, Spain}

\author{Massimiliano Stengel}
\affiliation{Institut de Ci\`encia de Materials de Barcelona 
(ICMAB-CSIC), Campus UAB, 08193 Bellaterra, Spain}
\affiliation{ICREA - Instituci\'o Catalana de Recerca i Estudis Avan\c{c}ats, 08010 Barcelona, Spain}

\date{\today}

\begin{abstract} 
Building on recent developments in electronic-structure methods, we 
define and calculate the flexoelectric response of two-dimensional (2D) materials fully from 
first principles.
In particular, we show that the open-circuit voltage response to a flexural deformation is a
fundamental linear-response property of the crystal that can be calculated
within the primitive unit cell of the flat configuration.
Applications to graphene, silicene, phosphorene, BN and transition-metal dichalcogenide monolayers
reveal that two distinct contributions exist, respectively of purely electronic and lattice-mediated
nature.
Within the former, we identify a key \emph{metric} term, consisting in the quadrupolar moment of the unperturbed 
charge density.
We propose a simple continuum model to connect our findings with the available experimental
measurements of the converse flexoelectric effect.
\end{abstract}

\maketitle

Among their many prospective applications, two-dimensional (2D) materials have 
received, in the last few years, considerable attention as a basis for novel
electromechanical device concepts, such as sensors or energy harvesters.~\cite{wu-14,ahmadpoor-15}
Such an interest has stimulated intense research, both experimental and theoretical,
to characterize the fundamentals of electromechanical couplings in monolayer (or few-layer)
graphene,~\cite{kalinin-08,mcgilly-20} boron nitride~\cite{naumov-09,duerloo-12} and transition-metal 
dichalcogenides.~\cite{brennan-17,wu-14}
For the most part, efforts were directed at understanding piezoelectric
and piezotronic properties~\cite{wu-14} with stretchable/tunable electronics
in mind; more recently flexoelectricity has been attracting increasing 
attention.~\cite{ahmadpoor-15,brennan-20}

Flexoelectricity, describing the coupling between a strain gradient and the 
macroscopic polarization,~\cite{zubko-13,wang-19} is expected to play a
prominent role in 2D crystals due to their extreme flexibility.
Recently, several experimental works~\cite{brennan-17,brennan-20,mcgilly-20} 
reported a significant out-of-plane electromechanical response in
graphene, BN, transition-metal dichalcogenides (TMDs) and related materials.
Experiments were generally performed via piezoelectric force microscopy (PFM), 
which probes the \emph{converse} effect (deformations in response to an applied 
voltage) in terms of an effective piezoelectric coefficient, $d^{\rm eff}_{33}$.
How the measured values of $d^{\rm eff}_{33}$ relate
to the intrinsic flexoelectric coefficients of the 2D layer is, however, 
currently unknown.
First, experiments are usually performed on supported layers;~\cite{brennan-17,brennan-20} 
this implies a suppression of their mechanical response due to substrate interaction,~\cite{suspended} 
whose impact on $d^{\rm eff}_{33}$ remains poorly understood. 
Second, flexoelectricity is a non-local 
effect, where electromechanical stresses depend on the \emph{gradients}
of the applied external field; this substantially complicates the analysis 
compared to the piezoelectric case, where spatial inhomogeneities in 
the tip potential play little role.~\cite{soergel-07} 
In fact, even understanding what components of the 2D flexoelectric tensor contribute
to $d^{\rm eff}_{33}$ is far from trivial.~\cite{brennan-17}
Unless these questions are settled by establishing reliable
models of the converse flexoelectric effect in 2D crystals, the analysis
of the experimental data remains to a large extent speculative, which
severely limits further progress towards a quantitative understanding.

Theoretical simulations are a natural choice to shed some light on the 
aforementioned issues. 
Several groups have studied flexoelectricity in a variety of 
monolayer crystals including graphene, hexagonal BN, and transition metal 
dichalcogenides; calculations were performed either from first 
principles~\cite{kalinin-08,naumov-09,shi-18,shi-19,pandey-21a,pandey-21b} or by means of
classical force fields.~\cite{zhuang-19,javvaji-19} 
Most authors, however, have defined and calculated
the flexoelectric coefficient as a dipolar moment of the deformed layer,
which has two main shortcomings. 
First, calculating the dipole moment 
of a curved crystalline slab is not free from ambiguities~\cite{codony-20}, and this has resulted in
a remarkable scattering of the reported results.
Second, such a definition has limited practical value, unless its relationship 
with the experimentally relevant parameters (electric fields and potentials) 
is established.
The latter issue may appear insignificant at first sight, but should not be
underestimated, as the Poisson equation of electrostatics is modified by 
curvature in a nontrivial way.~\cite{stengel-13b}
Some controversies around the thermodynamic equivalence between the direct
and converse flexoelectric effect~\cite{cross-06,yudin-13} complicate the
situation even further, calling for a fundamental solution to the problem.
Thanks to the progress of the past few years in the computational
methods,~\cite{resta-10,stengel-13a,stengel-13b,chapter-15,dreyer-18,schiaffino-19,royo-19} addressing these
questions in the framework of first-principles linear-response 
theory appears now well within reach.

Here we overcome the aforementioned limitations by defining and
calculating flexoelectricity as
the open-circuit voltage response to a flexural
deformation (``flexovoltage'') of the 2D crystal in the linear regime. 
Building on the recently-developed implementation of 
bulk flexoelectricity in 3D,~\cite{royo-19,romero-20} we show that 
the flexovoltage coefficient, $\varphi$, is a fundamental 
linear-response property of the crystal, and can be calculated 
by using the primitive 2D cell of the unperturbed flat layer.
We demonstrate our method by studying several monolayer materials as testcases
(C, Si, P, BN, MoS$_2$, WSe$_2$ and SnS$_2$), which we validate against direct
calculation of nanotube structures.
We find that the overall response consists in two well-defined contributions, 
a clamped-ion (CI) and a lattice-mediated (LM) term, in close analogy with the
theory of the piezoelectric response.~\cite{martin-72}
At the CI level, our calculations show a remarkable cancellation 
between a dipolar linear-response term and a previously overlooked ``metric''
contribution, which we rationalize in terms of an intuitive toy model
of noninteracting neutral spheres.~\cite{stengel-13b,chapter-15}
We further demonstrate that $\varphi$ describes both
the direct and converse coupling between local curvature and 
transverse electric fields in an arbitrary geometry, ranging
from nanotubes to flexural phonons and rippled layers. 
Based on this result, we build a quantitatively predictive
model of a flexoelectric layer on a substrate, and we use it to 
discuss recent experimental findings.

\begin{figure} 
\begin{center}
  \includegraphics[width=1.0\columnwidth]{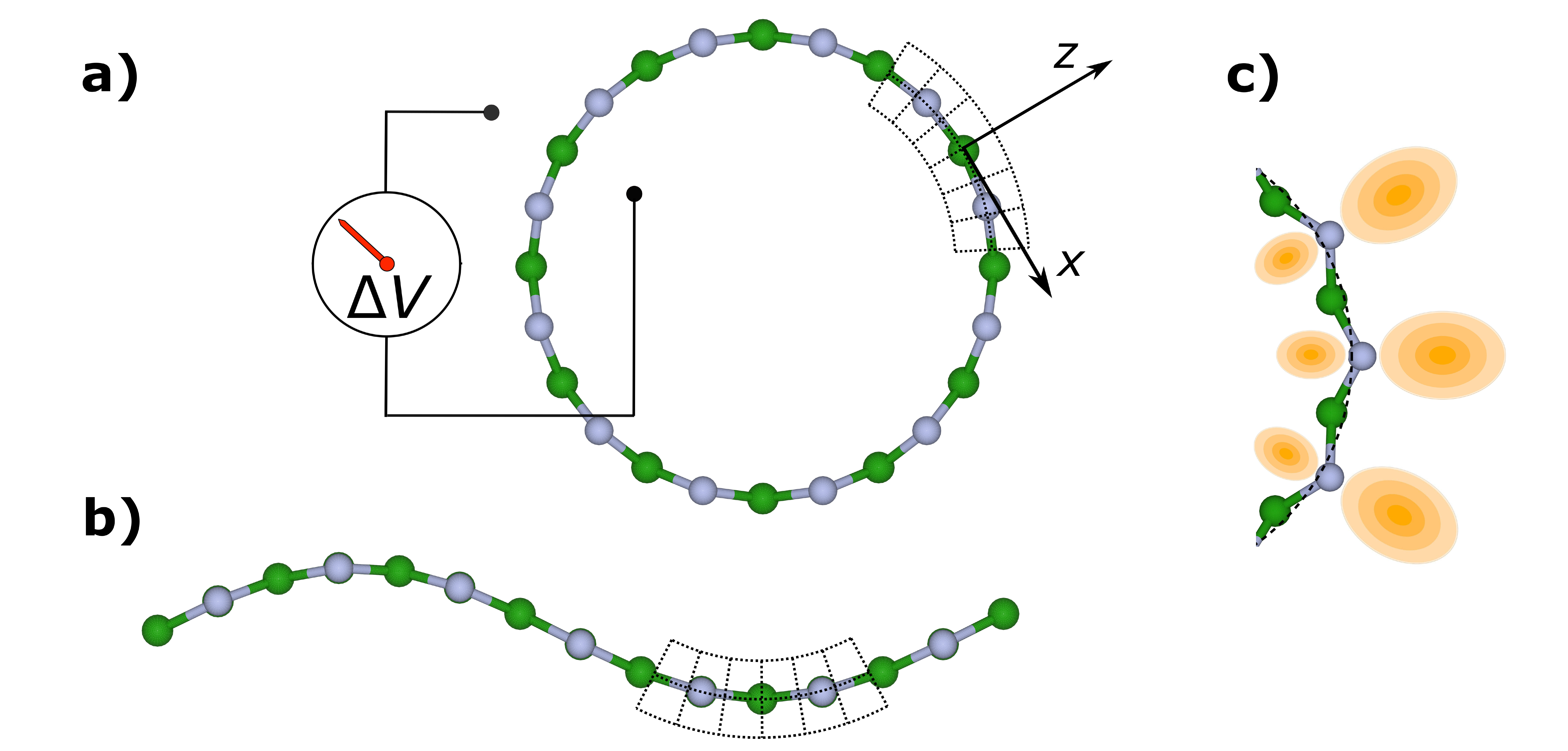}\\
  \caption{Schematic illustration of the 
  flexoelectric response in BN.
  (a): Cross-section of a BN nanotube; the voltage drop between its inner and outer sides is highlighted. (b): Flexural phonon, corresponding
  to an in-plane modulation of the same strain field as in (a). 
  (c): Lattice-mediated 
  and purely electronic effects contributing to the dipole. Gray/green circles represent the B/N atoms. The deformed electronic 
  orbitals are shown as yellow shaded ellipses.
  } 
\label{nanotube}
\end{center}
\end{figure}

The fundamental quantity that we shall address here is the voltage drop across a thin layer due to a 
flexural deformation, where the latter is measured by the radius of curvature, $R$.
At the leading order, the voltage drop is inversely proportional to $R$,
\begin{equation}
\Delta V = \dfrac{\varphi}{R} + O(R^{-2}), \qquad \varphi = \frac{\mu^{\rm 2D}}{\epsilon_0},
\label{Delta_V_1/R}
\end{equation} 
where $\varphi$ can also be expressed as a 2D flexoelectric coefficient (in units of charge,
describing the \emph{effective} dipole per unit area that is linearly induced by 
a flexural deformation) divided by the vacuum permittivity, $\epsilon_0$.
Our goal is to calculate the constant of proportionality, $\varphi$, which we shall
refer to as ``flexovoltage'' coefficient. 

The underlying physical model is that of a nanotube of radius $R$ constructed by bending a flat layer,
as illustrated in Fig.\ref{nanotube}(a); the voltage drop between the interior and the exterior is then 
given by Eq.~(\ref{Delta_V_1/R}).
Another obvious example is that of a long-wavelength
flexural phonon [see Fig.~\ref{nanotube}(b)]. 
Due to rotational invariance, the modulated strain field locally recovers 
the same pattern as in Fig.\ref{nanotube}(a). [See Ref.~\onlinecite{stengel-13b} and
Sec.~2.8.2 of Ref.~\onlinecite{chapter-15}.]
At the leading order in the wave vector, ${\bf q}$ this results in 
a local jump in the electrostatic potential across the layer of 
\begin{equation}
\label{kxy}
\Delta V (x,y) = \varphi K(x,y),
\end{equation} 
where the local inverse radius of curvature, $K(x,y) = -\nabla^2 u_z(x,y)$, 
is
given by the Laplacian of the vertical displacement 
field, $u_z$.~\cite{supplemental_prl} 
As we shall see shortly, the effect originates from the distortion of both the electronic cloud and
the crystal structure [Fig.\ref{nanotube}(c)].

To express the flexovoltage as a linear-response property, 
we start by associating the flexural deformation with a mapping between 
the Cartesian frame of the flat layer, and the curvilinear frame
of the bent nanotube [dashed grid in Fig.\ref{nanotube}(a)].
In a neighborhood of the nanotube surface, such a mapping 
corresponds to a strain field with 
cylindrical symmetry, of the type $\varepsilon_{xx}(z) = z/R$,
where $\hat{z}$ is the normal 
to the layer surface, $\hat{x}$ runs over the 
tangential direction, and $\varepsilon_{\alpha\beta}$ is
the symmetric strain tensor.
This results in a macroscopic transverse strain gradient 
$\varepsilon_{xx,z} = \partial \varepsilon_{xx} / \partial z = 1 / R,$
whose amplitude is the inverse radius of curvature,~\cite{stengel-14,chapter-15} 
and will play the role of the perturbative parameter, $\lambda$, henceforth.

To discuss the electrostatic potential, which defines 
the open-circuit voltage $\varphi$, we shall
frame our arguments on the Poisson equation in 
the curvilinear frame of the bent layer,
following the guidelines of Refs.~\onlinecite{stengel-13b,stengel-14,chapter-15},
\begin{equation}
\bm{\nabla} \cdot (\bm{\epsilon} \cdot {\bf E}) = \rho, \qquad \bm{\epsilon} = \epsilon_0 \sqrt{g} {\bf g}^{-1}.
\label{poisson}
\end{equation}
The main difference with respect to the Cartesian formulation is that 
the vacuum permittivity here becomes a tensor that depends on the metric of the 
deformation, ${\bf g}$. 
Within the linear regime, one can write $\rho = \rho^{(0)} + \lambda \rho^{(1)} + \cdots$,
where $\rho^{(0)}$ is the unperturbed density and $\rho^{(1)}$ the
first-order response in $\lambda$.
(For the time being we shall assume 
that $\rho^{(1)}$ refers to the static response, inclusive of electronic and
ionic relaxations.)
We find that the open-circuit potential $\varphi$ is given by
\begin{equation}
\varphi = \frac{\mathcal{D}[\rho^{(1)}]}{\epsilon_0}   + \varphi^{\rm M},
   \qquad     \varphi^{\rm M}=  
        - \dfrac{  \mathcal{Q}[ \rho^{(0)}] } {2 \epsilon_{0}},
  \label{phirho}
\end{equation}
where
\begin{equation}
 \mathcal{D}[f] =  \int dz \, z f(z)  \qquad \mathcal{Q}[f] =  \int dz \, z^{2} f(z).
\label{Qf}
\end{equation}
indicate the first (dipolar) and second (quadrupolar) moment of the function $f$ along 
the out-of-plane direction $z$, and $\rho^{(0)}(z)$ and $\rho^{(1)}(z)$ are the in-plane averages of
the respective microscopic response functions.
$\mathcal{D}[\rho^{(1)}]$ corresponds to the $\lambda$-derivative of the 
``radial polarization'' (${\bf p}$) as defined in Ref.~\onlinecite{codony-20}; 
the second term in Eq.~(\ref{phirho}) is a \emph{metric} contribution that only 
depends on the unperturbed density $\rho^{(0)}$, and originates from the linear 
variation of $\bm{\epsilon}$ in Eq.~(\ref{poisson}).
As we shall see shortly, the dipolar linear-response part is always large and negative, 
while the metric term is large and positive, typically leading to an
almost complete mutual cancellation.

The challenging part of the problem consists in computing
the dipolar linear-response contribution. 
To facilitate our progress towards a practical method, we shall use
$\rho^{(1)} = -\nabla \cdot {\bf P}^{(1)}$, where ${\bf P}^{(1)}$
is the microscopic \emph{polarization} response to the
deformation.
(The zero-th moment of $P^{(1)}_z$ along $z$ yields $\mathcal{D}[\rho^{(1)}]$,
after an integration by parts.)
Then, by using the formulation of Ref.~\cite{stengel-13b}, we can write the
radial component of ${\bf P}^{(1)}$ as
\begin{equation}
P^{(1)}_z(z) =
z P^{\rm U}_{z,xx}(z) + P^{\rm G}_{zz,xx}(z).
\label{pmic}
\end{equation}
The cell-periodic response functions $P^{\rm U}_{z,xx}(z)$ and 
$P^{\rm G}_{zz,xx}(z)$ (in-plane averaging is assumed) have the 
physical interpretation of a \emph{local} piezoelectric (U) and
flexoelectric (G) coefficient.~\cite{stengel-14}
The rationale behind such a decomposition is rooted on the 
availability of efficient first-principles methods to 
calculate both terms in Eq.~(\ref{pmic}), as we shall 
illustrate in the following.

To perform the actual calculations, we shall accommodate the 
unperturbed (flat) monolayer in a standard supercell, where
the out-of-plane dimension $L$ is treated as a convergence parameter.
Regarding the gradient (G) contribution, we find
\begin{equation}
\varphi^{\rm G} = \frac{1}{\epsilon_0} \int dz P^{\rm G}_{zz,xx}(z) = 
  \dfrac{L}{\epsilon_{0}\epsilon_{zz}}\mu_{zz,xx},
\label{phi_bf}
\end{equation}
where $\epsilon_{zz}$ is the out-of-plane component of the macroscopic 
dielectric tensor, and $\mu_{zz,xx}$ is the transverse component of the
flexoelectric tensor of the supercell.
Clearly, both $\mu_{zz,xx}$ and $\epsilon_{zz}$ depend on $L$ 
(averaging over an arbitrary supercell volume is implied, and short-circuit electrical
boundary conditions are usually imposed~\cite{royo-19} in the calculation of $\bm{\mu}$).
However, they do so in such a way that their ratio multiplied by $L$ does not (assuming 
that $L$ is large enough to consider the repeated layers as nonoverlapping). 
Regarding the contribution  
of the first term on the rhs of Eq.~(\ref{pmic}),
we have  
\begin{equation}
\varphi^{\rm U}= \frac{ \mathcal{D}[P^{\rm U}_{z,xx}]} {\epsilon_{0}}=
\dfrac{ \mathcal{Q}[ \rho^{\rm U}_{xx}]} {2 \epsilon_{0}}, 
\label{phi_sp} 
\end{equation}
where $\rho^{\rm U}_{xx}(z)$ is the
first-order charge-density response to a uniform
strain ($\rho^{\rm U}_{xx} = -\partial P^{\rm U}_{z,xx} /\partial z$). 
The total flexovoltage of the slab is then given by 
\begin{equation}
\varphi = \frac{dV}{d \lambda} = 
  \varphi^{\rm G} + \varphi^{\rm U}
   +  \varphi^{\rm M},
\label{phi_tot}
\end{equation}
where neither of $\varphi^{\rm G}$, $\varphi^{\rm U}$ or $\varphi^{\rm M}$ depend on $L$,
and should therefore be regarded as well-defined physical properties of the isolated
monolayer.
One can verify that, by applying the present formulation to crystalline slabs of 
increasing thickness, we recover the results of Ref.~\onlinecite{stengel-14} once $\varphi$ is 
divided by the slab thickness, $t$, and the thermodynamic limit performed. 
($\varphi^{\rm G}$ and $\varphi^{\rm U} + \varphi^{\rm M}$ tend to the bulk and surface 
contributions to the total flexoelectric effect, respectively.)

Eq.~(\ref{phi_tot}) is directly suitable for a numerical implementation, as it only requires 
response functions that are routinely calculated within density-functional perturbation theory (DFPT).
The partition between ``G'' and ``U'' contributions, however, is hardly meaningful for an
atomically thin 2D monolayer, where essentially everything is surface and there is no bulk
underneath.
Thus, we shall recast Eq.~(\ref{phi_tot}) in a more useful form hereafter, by
seeking a separation between clamped-ion (CI) and lattice-mediated (LM) effects 
instead [Fig.~\ref{nanotube}(b)], 
\begin{equation}
\varphi = \varphi^{\rm CI} + \varphi^{\rm LM}.
\label{phi_tot2}
\end{equation}

We find~\cite{supplemental_prl} that the CI contribution
has the same functional form as the total response, 
\begin{eqnarray}
\varphi^{\rm CI} &=& \dfrac{L}{\epsilon_{0} \bar{\epsilon}_{zz}} \bar{\mu}_{zz,xx}
 + \dfrac{ \mathcal{Q}[ \bar{\rho}^{\rm U}_{xx}] } {2 \epsilon_{0}}-  
   \dfrac{      \mathcal{Q}[ \rho^{(0)}] } {2 \epsilon_{0}}, \label{phi_CI}
\end{eqnarray}
with the only difference that the flexoelectric ($\mu$), dielectric ($\epsilon_{zz}$) and uniform-strain charge
response ($\rho^{\rm U}$) functions have been replaced here with their clamped-ion counterparts, indicated by
barred symbols. 
Regarding the LM part, 
\begin{eqnarray}
\varphi^{\rm LM} &=&  \frac{1}{S \epsilon_0} \hat{Z}^{(z)}_{\kappa \alpha} \, \hat{\Phi}^{-1}_{\kappa \alpha, \kappa' \beta}
 \, \hat{\mathcal{C}}^{\kappa'}_{\beta z,xx},
\label{phi_LM}
\end{eqnarray}
we have a more intuitive description in terms of the out-of-plane \emph{longitudinal}
charges $\hat{Z}^{(z)}_{\kappa \alpha} = {Z}^{(z)}_{\kappa \alpha} / \bar{\epsilon}_{zz}$, the
pseudoinverse~\cite{hong-13} of the zone-center dynamical matrix, $\hat{\Phi}^{-1}_{\kappa \alpha, \kappa' \beta}$, and the  
atomic force response~\cite{supplemental_prl} to a flexural deformation of the slab, $\hat{\mathcal{C}}^{\kappa}_{\beta z,xx}$.
Note that the ``mixed'' contribution~\cite{stengel-13a} to the bulk flexoelectric tensor exactly
cancels~\cite{supplemental_prl} with an equal and opposite term in the lattice-mediated contribution 
to $\varphi^{\rm U}$, hence its absence from Eq.~(\ref{phi_tot2}).

\begin{table} 
\setlength{\tabcolsep}{10pt}
\begin{center}
\begin{tabular}{c|ccc}\hline\hline
  \T\B  & \multicolumn{1}{c}{$\varphi^{\rm CI}$}  & $\varphi^{\rm LM}$ & $\varphi$  \\\hline
C       \T\B &           $-$0.1134 & \hspace{4pt} 0.0000 &           $-$0.1134  \\  
Si      \T\B & \hspace{4pt} 0.0585 & \hspace{4pt} 0.0000 & \hspace{4pt} 0.0585  \\
P (zigzag)  \T\B & \hspace{4pt} 0.2320 &           $-$0.0151 & \hspace{4pt} 0.2170 \\ 
P (armchair) \T\B & $-$0.0130 & $-$0.0461 & $-$0.0591 \\
BN      \T\B & $-$0.0381 & $-$0.1628 & $-$0.2009  \\
MoS$_2$ \T\B & $-$0.2704 & $-$0.0565 & $-$0.3269  \\
WSe$_2$ \T\B & $-$0.3158 & $-$0.0742 & $-$0.3899  \\ 
SnS$_2$ \T\B & \hspace{4pt} 0.1864 & \hspace{4pt} 0.1728 & \hspace{4pt} 0.3592\\ \hline \hline
\end{tabular}
\end{center}
\caption{Clamped-ion (CI), lattice-mediated (LM) and total flexovoltages (nV$\cdot$m) of 
the 2D crystals studied in this work. Due to its lower symmetry, for phosphorene two independent 
bending directions (armchair and zigzag) exist.} 
\label{Rel_zzxx_flexo}
\end{table}

Our calculations are performed in the framework of density-functional perturbation theory~\cite{baroni-01,gonze-97b} 
(DFPT) within the local-density approximation, 
as implemented in ABINIT~\cite{abinit,romero-20}. 
(Computational parameters and extensive tests, including calculations 
performed within the generalized-gradient approximation, 
are described in Ref.~\onlinecite{supplemental_prl}).
In Table \ref{Rel_zzxx_flexo} we report the 
calculated bending flexovoltages 
for several monolayer crystals. 
Both the CI and LM contributions 
show a considerable variety in magnitude 
and sign: while the former dominates in the TMDs, the
reverse is true for BN, and SnS$_2$ seems to lie 
right in the middle.
The case of phosphorene is interesting: its lower
symmetry allows for a nonzero $\varphi^{\rm LM}$ in
spite of it being an elemental crystal like C and Si; 
it also allows for a substantial anisotropy of the response.
If we assume a physical thickness $t$ corresponding to the bulk interlayer spacing, 
we obtain an estimate (see Table 6 of Ref.~\onlinecite{supplemental_prl}) for the volume-averaged flexoelectric coefficients, 
of $|\mu| = |\mu^{\rm 2D}|/t \sim 1-5$ pC/m.
($\mu$, unlike $\mu^{\rm 2D}$, is inappropriate~\cite{codony-20} 
for 2D layers given the ill-defined nature of the parameter $t$;
we use it here for comparison purposes only.)
This value is in the same ballpark as earlier predictions,~\cite{shi-18,zhuang-19,pandey-21a,pandey-21b} 
although there is a considerable scatter in the latter.
For example, the value quoted by Ref.~\cite{zhuang-19} for graphene is very close 
to ours, but their results for other materials are either much larger (TMD's, silicene) 
or much smaller (BN); other works tend to disagree both with our results and among 
themselves.
These large discrepancies are likely due to the specific computational methods that were 
adopted in each case (often the total dipole moment of a bent nanoribbon including the boundaries
was calculated, rather than the intrinsic response of the extended layer), 
or to the aforementioned difficulties~\cite{codony-20} with 
the definition of the dipole of a curved surface.

Very recently Ref.~\cite{codony-20} reported first-principles 
calculations of some of the materials presented here by using 
methods that bear some similarities to ours, which allows for a more 
meaningful comparison.
By converting our results for Si and C to the units of Ref.~\cite{codony-20} 
via Eq.~(\ref{Delta_V_1/R}), we obtain $\mu_{\rm C} = -0.0063e$ and 
$\mu_{\rm Si} = +0.0032e$; these, however, are almost two orders of magnitude
smaller, and with inconsistent signs, with respect to the corresponding
results of Ref.~\cite{codony-20}.
We ascribe the source of disagreement to the neglect in Ref.~\cite{codony-20} 
of the metric term in Eq.~(\ref{phi_CI}). 
Indeed, for the dipolar linear-response contribution [first two terms in Eq.~(\ref{phi_CI})] 
we obtain $\mu^{\rm dip}_{\rm C} = -0.22e$ and  $\mu^{\rm dip}_{\rm Si} = -0.19e$,
now in excellent agreement (except for the sign) with the results of Codony {\em et al.}.
This observation points to a nearly complete cancellation between the dipolar and
metric contribution to $\varphi$, which is systematic across the whole materials set
(see Table 3 of Ref.~\cite{supplemental_prl}).

To clarify this point,
we have performed additional calculations on toy model, 
consisting of a hexagonal layer of well-spaced rare gas atoms.~\cite{supplemental_prl}
This is a system where no response should occur, as 
an arbitrary ``mechanical deformation'' consists
in the trivial displacement of noninteracting (and spherically 
symmetric) neutral atoms.
We find that $\varphi^{\rm dip}$ and $\varphi^{\rm met}$ are,
like in other cases, large and opposite in sign; this, however, 
is just a side-effect of the coordinate transformation (i.e., a 
mathematical artefact), and does not reflect a true physical response
of the system to the perturbation.
For a lattice parameter that is large enough, 
the cancellation becomes exact and 
our calculated value of $\varphi$ vanishes as expected on 
physical grounds.
This further corroborates the soundness of our definition of
$\mu^{\rm 2D}$, which is based on the electrostatic
potential. 
The latter, in addition to being an experimentally relevant parameter,
behaves as a true scalar under a coordinate transformation; it is 
therefore unaffected, unlike the charge density, by the 
(arbitrary) 
choice of the reference frame.~\cite{supplemental_prl}

\begin{figure} 
\begin{center}
  \includegraphics[width=0.9\columnwidth]{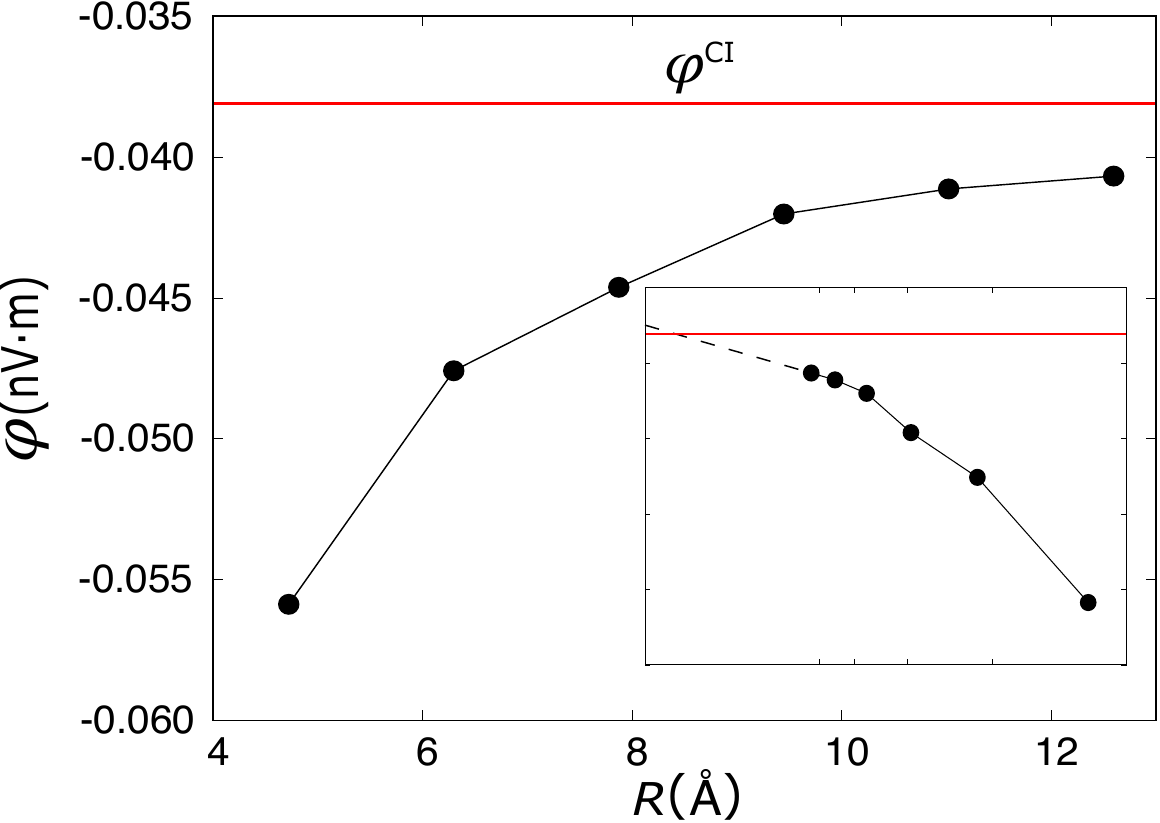}\\
  \caption{Clamped-ion flexovoltage coefficient calculated as $\varphi =  \Delta V \, R$, plotted as a function of the nanotube radius $R$.
  Our linear-response result [Eq.~(\ref{phi_CI})] is shown as a red line. The inset shows $\varphi$ as
  a function of $1/R$, the dashed line being a linear extrapolation from the last two calculated points 
  to $R\rightarrow\infty$.} 
  \label{fig_nt}
\end{center}
\end{figure}

As a further consistency check, we have performed explicit calculations of
BN nanotubes of increasing radius $R$, and extracted the voltage drop between their interior and
exterior, $\Delta V$, 
at the clamped-ion level.
In Figure \ref{fig_nt} we plot the estimated flexovoltage, given by $R \, \Delta V$, 
as a function of $R$. 
The asymptotic convergence 
to the linear-response value of
$\varphi^{\rm CI}$ 
is clear, consistent with Eq.~\eqref{Delta_V_1/R}.
The convergence rate, however, appears rather slow: at the 
largest value of $R$, corresponding to a nanotube primitive cell of 128 atoms, the deviation from 
$\varphi^{\rm CI}$ is still of about 10\%.
This result highlights the difficulties at calculating flexovoltages in 2D systems by using 
the direct approach; 
conversely, our method provides an optimally converged solution within few minutes on a modern 
workstation, and is ideally suited, e.g. for high-throughput screening applications.

The implications of our findings for the interpretation of the experiments
are best discussed in terms of the interaction between the flexural modes of a 
flat layer and an external, generally inhomogeneous, out-of-plane electric 
field, $\mathcal{E}_z(x,y)$.
In full generality, Eq.~(\ref{kxy}) leads to the following
coupling (energy per unit area),
\begin{equation}
\label{efl}
E_{\rm flexo}(u_z,\mathcal{E}_z) = \mu^{\rm 2D} \mathcal{E}_z \, \nabla^2 u_z,
\end{equation}
which reduces to $E_{\rm flexo} = -q^2 \mu^{\rm 2D} \mathcal{E}_z u_z$ for
monochromatic fields of the type $A(x,y) = A \cos({\bf q\cdot r})$ [$A=(u_z,\mathcal{E}_z)]$.
By deriving $E_{\rm flexo}$ with respect to the displacement $u_z$ we obtain the \emph{converse}
flexoelectric effect, in the form of a vertical force per unit area, $\mathcal{F}_z = q^2
\mu^{\rm 2D} \mathcal{E}_z$, in response to the field.
Explicit first-principles calculations of a BN layer under an applied 
$\mathcal{E}_z$~\cite{supplemental_prl} nicely confirm this prediction: Eq.~(\ref{efl}) is
the main source of out-of-plane electromechanical response 
in this class of materials.
Note that the \emph{longitudinal} out-of-plane flexoelectric coefficient of a 
free-standing layer, which we extract as a by-product of our main calculations, always
vanishes (see Ref.~\onlinecite{supplemental_prl}) due to translational invariance 
and thus cannot contribute to the coupling, contrary to
the common belief.~\cite{brennan-17,brennan-20}.

This allows us to generalize the existing models~\cite{amorim-13}
of supported 2D layers by incorporating flexoelectricity, and thereby
extract two important messages.~\cite{supplemental_prl}
First, the amplitude of the response is highly sensitive to the substrate interaction
strength, $g$, consistent with the results of recent measurements performed on suspended 
layers.~\cite{suspended} 
Second, the response displays a strong dispersion in $q$, indicating a
marked sensitivity on the length scale of the inhomogeneities in the
applied field.
Both outcomes call for a reinterpretation of the existing PFM
measurements of flexoelectricity:~\cite{brennan-17} information
about $g$ and the tip geometry appears essential for a quantitative
estimation of $\mu^{\rm 2D}$. 
We hope that our results will stimulate further experimental research 
along these lines, and more generally to facilitate 
the design of piezoelectric nanocomposites~\cite{chu-09} based on the flexoelectric effect. 

\nocite{royo-20,HamannPRB13,pseudodojo,novoselov-16,pike-17,pbe,hamann_gga,Jain-13,screen_2D,kumar-20,screen_2D}

\begin{acknowledgments}
 We acknowledge the support of Ministerio de Economia,
 Industria y Competitividad (MINECO-Spain) through
 Grants  No.  MAT2016-77100-C2-2-P  and  No.  SEV-2015-0496,
 and  of Generalitat de Catalunya (Grant No. 2017 SGR1506).
 This project has received funding from the European
 Research Council (ERC) under the European Union's
 Horizon 2020 research and innovation program (Grant
 Agreement No. 724529). Part of the calculations were performed at
 the Supercomputing Center of Galicia (CESGA).
\end{acknowledgments}

\bibliography{flexo2d}

\begin{thebibliography}{48}%
\makeatletter
\providecommand \@ifxundefined [1]{%
 \@ifx{#1\undefined}
}%
\providecommand \@ifnum [1]{%
 \ifnum #1\expandafter \@firstoftwo
 \else \expandafter \@secondoftwo
 \fi
}%
\providecommand \@ifx [1]{%
 \ifx #1\expandafter \@firstoftwo
 \else \expandafter \@secondoftwo
 \fi
}%
\providecommand \natexlab [1]{#1}%
\providecommand \enquote  [1]{``#1''}%
\providecommand \bibnamefont  [1]{#1}%
\providecommand \bibfnamefont [1]{#1}%
\providecommand \citenamefont [1]{#1}%
\providecommand \href@noop [0]{\@secondoftwo}%
\providecommand \href [0]{\begingroup \@sanitize@url \@href}%
\providecommand \@href[1]{\@@startlink{#1}\@@href}%
\providecommand \@@href[1]{\endgroup#1\@@endlink}%
\providecommand \@sanitize@url [0]{\catcode `\\12\catcode `\$12\catcode
  `\&12\catcode `\#12\catcode `\^12\catcode `\_12\catcode `\%12\relax}%
\providecommand \@@startlink[1]{}%
\providecommand \@@endlink[0]{}%
\providecommand \url  [0]{\begingroup\@sanitize@url \@url }%
\providecommand \@url [1]{\endgroup\@href {#1}{\urlprefix }}%
\providecommand \urlprefix  [0]{URL }%
\providecommand \Eprint [0]{\href }%
\providecommand \doibase [0]{http://dx.doi.org/}%
\providecommand \selectlanguage [0]{\@gobble}%
\providecommand \bibinfo  [0]{\@secondoftwo}%
\providecommand \bibfield  [0]{\@secondoftwo}%
\providecommand \translation [1]{[#1]}%
\providecommand \BibitemOpen [0]{}%
\providecommand \bibitemStop [0]{}%
\providecommand \bibitemNoStop [0]{.\EOS\space}%
\providecommand \EOS [0]{\spacefactor3000\relax}%
\providecommand \BibitemShut  [1]{\csname bibitem#1\endcsname}%
\let\auto@bib@innerbib\@empty
\bibitem [{\citenamefont {Wu}\ \emph {et~al.}(2014)\citenamefont {Wu},
  \citenamefont {Wang}, \citenamefont {Li}, \citenamefont {Zhang},
  \citenamefont {Lin}, \citenamefont {Niu}, \citenamefont {Chenet},
  \citenamefont {Zhang}, \citenamefont {Hao}, \citenamefont {Heinz},
  \citenamefont {Hone},\ and\ \citenamefont {Wang}}]{wu-14}%
  \BibitemOpen
  \bibfield  {author} {\bibinfo {author} {\bibfnamefont {Wenzhuo}\ \bibnamefont
  {Wu}}, \bibinfo {author} {\bibfnamefont {Lei}\ \bibnamefont {Wang}}, \bibinfo
  {author} {\bibfnamefont {Yilei}\ \bibnamefont {Li}}, \bibinfo {author}
  {\bibfnamefont {Fan}\ \bibnamefont {Zhang}}, \bibinfo {author} {\bibfnamefont
  {Long}\ \bibnamefont {Lin}}, \bibinfo {author} {\bibfnamefont {Simiao}\
  \bibnamefont {Niu}}, \bibinfo {author} {\bibfnamefont {Daniel}\ \bibnamefont
  {Chenet}}, \bibinfo {author} {\bibfnamefont {Xian}\ \bibnamefont {Zhang}},
  \bibinfo {author} {\bibfnamefont {Yufeng}\ \bibnamefont {Hao}}, \bibinfo
  {author} {\bibfnamefont {Tony~F.}\ \bibnamefont {Heinz}}, \bibinfo {author}
  {\bibfnamefont {James}\ \bibnamefont {Hone}}, \ and\ \bibinfo {author}
  {\bibfnamefont {Zhong~Lin}\ \bibnamefont {Wang}},\ }\bibfield  {title}
  {\enquote {\bibinfo {title} {Piezoelectricity of single-atomic-layer
  {MoS$_2$} for energy conversion and piezotronics},}\ }\href@noop {}
  {\bibfield  {journal} {\bibinfo  {journal} {Nature}\ }\textbf {\bibinfo
  {volume} {514}},\ \bibinfo {pages} {470--474} (\bibinfo {year}
  {2014})}\BibitemShut {NoStop}%
\bibitem [{\citenamefont {Ahmadpoor}\ and\ \citenamefont
  {Sharma}(2015)}]{ahmadpoor-15}%
  \BibitemOpen
  \bibfield  {author} {\bibinfo {author} {\bibfnamefont {Fatemeh}\ \bibnamefont
  {Ahmadpoor}}\ and\ \bibinfo {author} {\bibfnamefont {Pradeep}\ \bibnamefont
  {Sharma}},\ }\bibfield  {title} {\enquote {\bibinfo {title} {Flexoelectricity
  in two-dimensional crystalline and biological membranes},}\ }\href {\doibase
  10.1039/C5NR04722F} {\bibfield  {journal} {\bibinfo  {journal} {Nanoscale}\
  }\textbf {\bibinfo {volume} {7}},\ \bibinfo {pages} {16555--16570} (\bibinfo
  {year} {2015})}\BibitemShut {NoStop}%
\bibitem [{\citenamefont {Kalinin}\ and\ \citenamefont
  {Meunier}(2008)}]{kalinin-08}%
  \BibitemOpen
  \bibfield  {author} {\bibinfo {author} {\bibfnamefont {Sergei~V.}\
  \bibnamefont {Kalinin}}\ and\ \bibinfo {author} {\bibfnamefont {Vincent}\
  \bibnamefont {Meunier}},\ }\bibfield  {title} {\enquote {\bibinfo {title}
  {Electronic flexoelectricity in low-dimensional systems},}\ }\href {\doibase
  10.1103/PhysRevB.77.033403} {\bibfield  {journal} {\bibinfo  {journal} {Phys.
  Rev. B}\ }\textbf {\bibinfo {volume} {77}},\ \bibinfo {pages} {033403}
  (\bibinfo {year} {2008})}\BibitemShut {NoStop}%
\bibitem [{\citenamefont {McGilly}\ \emph {et~al.}(2020)\citenamefont
  {McGilly}, \citenamefont {Kerelsky}, \citenamefont {Finney}, \citenamefont
  {Shapovalov}, \citenamefont {Shih}, \citenamefont {Ghiotto}, \citenamefont
  {Zeng}, \citenamefont {Moore}, \citenamefont {Wu}, \citenamefont {Bai},
  \citenamefont {Watanabe}, \citenamefont {Taniguchi}, \citenamefont {Stengel},
  \citenamefont {Zhou}, \citenamefont {Hone}, \citenamefont {Zhu},
  \citenamefont {Basov}, \citenamefont {Dean}, \citenamefont {Dreyer},\ and\
  \citenamefont {Pasupathy}}]{mcgilly-20}%
  \BibitemOpen
  \bibfield  {author} {\bibinfo {author} {\bibfnamefont {Leo~J.}\ \bibnamefont
  {McGilly}}, \bibinfo {author} {\bibfnamefont {Alexander}\ \bibnamefont
  {Kerelsky}}, \bibinfo {author} {\bibfnamefont {Nathan~R.}\ \bibnamefont
  {Finney}}, \bibinfo {author} {\bibfnamefont {Konstantin}\ \bibnamefont
  {Shapovalov}}, \bibinfo {author} {\bibfnamefont {En-Min}\ \bibnamefont
  {Shih}}, \bibinfo {author} {\bibfnamefont {Augusto}\ \bibnamefont {Ghiotto}},
  \bibinfo {author} {\bibfnamefont {Yihang}\ \bibnamefont {Zeng}}, \bibinfo
  {author} {\bibfnamefont {Samuel~L.}\ \bibnamefont {Moore}}, \bibinfo {author}
  {\bibfnamefont {Wenjing}\ \bibnamefont {Wu}}, \bibinfo {author}
  {\bibfnamefont {Yusong}\ \bibnamefont {Bai}}, \bibinfo {author}
  {\bibfnamefont {Kenji}\ \bibnamefont {Watanabe}}, \bibinfo {author}
  {\bibfnamefont {Takashi}\ \bibnamefont {Taniguchi}}, \bibinfo {author}
  {\bibfnamefont {Massimiliano}\ \bibnamefont {Stengel}}, \bibinfo {author}
  {\bibfnamefont {Lin}\ \bibnamefont {Zhou}}, \bibinfo {author} {\bibfnamefont
  {James}\ \bibnamefont {Hone}}, \bibinfo {author} {\bibfnamefont {Xiaoyang}\
  \bibnamefont {Zhu}}, \bibinfo {author} {\bibfnamefont {Dmitri~N.}\
  \bibnamefont {Basov}}, \bibinfo {author} {\bibfnamefont {Cory}\ \bibnamefont
  {Dean}}, \bibinfo {author} {\bibfnamefont {Cyrus~E.}\ \bibnamefont {Dreyer}},
  \ and\ \bibinfo {author} {\bibfnamefont {Abhay~N.}\ \bibnamefont
  {Pasupathy}},\ }\bibfield  {title} {\enquote {\bibinfo {title} {Visualization
  of {Moir{\'e}} superlattices},}\ }\href@noop {} {\bibfield  {journal}
  {\bibinfo  {journal} {Nature Nanotechnology}\ }\textbf {\bibinfo {volume}
  {15}},\ \bibinfo {pages} {580--584} (\bibinfo {year} {2020})}\BibitemShut
  {NoStop}%
\bibitem [{\citenamefont {Naumov}\ \emph {et~al.}(2009)\citenamefont {Naumov},
  \citenamefont {Bratkovsky},\ and\ \citenamefont {Ranjan}}]{naumov-09}%
  \BibitemOpen
  \bibfield  {author} {\bibinfo {author} {\bibfnamefont {Ivan}\ \bibnamefont
  {Naumov}}, \bibinfo {author} {\bibfnamefont {Alexander~M.}\ \bibnamefont
  {Bratkovsky}}, \ and\ \bibinfo {author} {\bibfnamefont {V.}~\bibnamefont
  {Ranjan}},\ }\bibfield  {title} {\enquote {\bibinfo {title} {Unusual
  flexoelectric effect in two-dimensional noncentrosymmetric $s{p}^{2}$-bonded
  crystals},}\ }\href {\doibase 10.1103/PhysRevLett.102.217601} {\bibfield
  {journal} {\bibinfo  {journal} {Phys. Rev. Lett.}\ }\textbf {\bibinfo
  {volume} {102}},\ \bibinfo {pages} {217601} (\bibinfo {year}
  {2009})}\BibitemShut {NoStop}%
\bibitem [{\citenamefont {Duerloo}\ \emph {et~al.}(2012)\citenamefont
  {Duerloo}, \citenamefont {Ong},\ and\ \citenamefont {Reed}}]{duerloo-12}%
  \BibitemOpen
  \bibfield  {author} {\bibinfo {author} {\bibfnamefont {Karel-Alexander~N.}\
  \bibnamefont {Duerloo}}, \bibinfo {author} {\bibfnamefont {Mitchell~T.}\
  \bibnamefont {Ong}}, \ and\ \bibinfo {author} {\bibfnamefont {Evan~J.}\
  \bibnamefont {Reed}},\ }\bibfield  {title} {\enquote {\bibinfo {title}
  {Intrinsic piezoelectricity in two-dimensional materials},}\ }\href@noop {}
  {\bibfield  {journal} {\bibinfo  {journal} {The Journal of Physical Chemistry
  Letters}\ }\textbf {\bibinfo {volume} {3}},\ \bibinfo {pages} {2871--2876}
  (\bibinfo {year} {2012})}\BibitemShut {NoStop}%
\bibitem [{\citenamefont {Brennan}\ \emph {et~al.}(2017)\citenamefont
  {Brennan}, \citenamefont {Ghosh}, \citenamefont {Koul}, \citenamefont
  {Banerjee}, \citenamefont {Lu},\ and\ \citenamefont {Yu}}]{brennan-17}%
  \BibitemOpen
  \bibfield  {author} {\bibinfo {author} {\bibfnamefont {Christopher~J.}\
  \bibnamefont {Brennan}}, \bibinfo {author} {\bibfnamefont {Rudresh}\
  \bibnamefont {Ghosh}}, \bibinfo {author} {\bibfnamefont {Kalhan}\
  \bibnamefont {Koul}}, \bibinfo {author} {\bibfnamefont {Sanjay~K.}\
  \bibnamefont {Banerjee}}, \bibinfo {author} {\bibfnamefont {Nanshu}\
  \bibnamefont {Lu}}, \ and\ \bibinfo {author} {\bibfnamefont {Edward~T.}\
  \bibnamefont {Yu}},\ }\bibfield  {title} {\enquote {\bibinfo {title}
  {Out-of-plane electromechanical response of monolayer molybdenum disulfide
  measured by piezoresponse force microscopy},}\ }\href@noop {} {\bibfield
  {journal} {\bibinfo  {journal} {Nano Letters}\ }\textbf {\bibinfo {volume}
  {17}},\ \bibinfo {pages} {5464--5471} (\bibinfo {year} {2017})}\BibitemShut
  {NoStop}%
\bibitem [{\citenamefont {Brennan}\ \emph {et~al.}(2020)\citenamefont
  {Brennan}, \citenamefont {Koul}, \citenamefont {Lu},\ and\ \citenamefont
  {Yu}}]{brennan-20}%
  \BibitemOpen
  \bibfield  {author} {\bibinfo {author} {\bibfnamefont {Christopher~J.}\
  \bibnamefont {Brennan}}, \bibinfo {author} {\bibfnamefont {Kalhan}\
  \bibnamefont {Koul}}, \bibinfo {author} {\bibfnamefont {Nanshu}\ \bibnamefont
  {Lu}}, \ and\ \bibinfo {author} {\bibfnamefont {Edward~T.}\ \bibnamefont
  {Yu}},\ }\bibfield  {title} {\enquote {\bibinfo {title} {Out-of-plane
  electromechanical coupling in transition metal dichalcogenides},}\ }\href
  {\doibase 10.1063/1.5134091} {\bibfield  {journal} {\bibinfo  {journal}
  {Applied Physics Letters}\ }\textbf {\bibinfo {volume} {116}},\ \bibinfo
  {pages} {053101} (\bibinfo {year} {2020})}\BibitemShut {NoStop}%
\bibitem [{\citenamefont {Zubko}\ \emph {et~al.}(2013)\citenamefont {Zubko},
  \citenamefont {Catalan},\ and\ \citenamefont {Tagantsev}}]{zubko-13}%
  \BibitemOpen
  \bibfield  {author} {\bibinfo {author} {\bibfnamefont {P.}~\bibnamefont
  {Zubko}}, \bibinfo {author} {\bibfnamefont {G.}~\bibnamefont {Catalan}}, \
  and\ \bibinfo {author} {\bibfnamefont {A.~K.}\ \bibnamefont {Tagantsev}},\
  }\bibfield  {title} {\enquote {\bibinfo {title} {Flexoelectric effect in
  solids},}\ }\href@noop {} {\bibfield  {journal} {\bibinfo  {journal} {Annu.
  Rev. Mater. Res.}\ }\textbf {\bibinfo {volume} {43}},\ \bibinfo {pages}
  {387--421} (\bibinfo {year} {2013})}\BibitemShut {NoStop}%
\bibitem [{\citenamefont {Wang}\ \emph
  {et~al.}(2019{\natexlab{a}})\citenamefont {Wang}, \citenamefont {Gu},
  \citenamefont {Zhang},\ and\ \citenamefont {Chen}}]{wang-19}%
  \BibitemOpen
  \bibfield  {author} {\bibinfo {author} {\bibfnamefont {Bo}~\bibnamefont
  {Wang}}, \bibinfo {author} {\bibfnamefont {Yijia}\ \bibnamefont {Gu}},
  \bibinfo {author} {\bibfnamefont {Shujun}\ \bibnamefont {Zhang}}, \ and\
  \bibinfo {author} {\bibfnamefont {Long-Qing}\ \bibnamefont {Chen}},\
  }\bibfield  {title} {\enquote {\bibinfo {title} {Flexoelectricity in solids:
  Progress, challenges, and perspectives},}\ }\href {\doibase
  10.1016/j.pmatsci.2019.05.003} {\bibfield  {journal} {\bibinfo  {journal}
  {Progress in Materials Science}\ }\textbf {\bibinfo {volume} {106}},\
  \bibinfo {pages} {100570} (\bibinfo {year} {2019}{\natexlab{a}})}\BibitemShut
  {NoStop}%
\bibitem [{\citenamefont {Wang}\ \emph
  {et~al.}(2019{\natexlab{b}})\citenamefont {Wang}, \citenamefont {Cui},
  \citenamefont {Chen}, \citenamefont {Xu}, \citenamefont {Hu}, \citenamefont
  {Jiang}, \citenamefont {Shang},\ and\ \citenamefont {Chu}}]{suspended}%
  \BibitemOpen
  \bibfield  {author} {\bibinfo {author} {\bibfnamefont {Xiang}\ \bibnamefont
  {Wang}}, \bibinfo {author} {\bibfnamefont {Anyang}\ \bibnamefont {Cui}},
  \bibinfo {author} {\bibfnamefont {Fangfang}\ \bibnamefont {Chen}}, \bibinfo
  {author} {\bibfnamefont {Liping}\ \bibnamefont {Xu}}, \bibinfo {author}
  {\bibfnamefont {Zhigao}\ \bibnamefont {Hu}}, \bibinfo {author} {\bibfnamefont
  {Kai}\ \bibnamefont {Jiang}}, \bibinfo {author} {\bibfnamefont {Liyan}\
  \bibnamefont {Shang}}, \ and\ \bibinfo {author} {\bibfnamefont {Junhao}\
  \bibnamefont {Chu}},\ }\bibfield  {title} {\enquote {\bibinfo {title}
  {Probing effective out-of-plane piezoelectricity in van der waals layered
  materials induced by flexoelectricity},}\ }\href {\doibase
  10.1002/smll.201903106} {\bibfield  {journal} {\bibinfo  {journal} {Small}\
  }\textbf {\bibinfo {volume} {15}},\ \bibinfo {pages} {1903106} (\bibinfo
  {year} {2019}{\natexlab{b}})}\BibitemShut {NoStop}%
\bibitem [{\citenamefont {Jungk}\ \emph {et~al.}(2007)\citenamefont {Jungk},
  \citenamefont {Hoffmann},\ and\ \citenamefont {Soergel}}]{soergel-07}%
  \BibitemOpen
  \bibfield  {author} {\bibinfo {author} {\bibfnamefont {T.}~\bibnamefont
  {Jungk}}, \bibinfo {author} {\bibfnamefont {{\'A}.}~\bibnamefont {Hoffmann}},
  \ and\ \bibinfo {author} {\bibfnamefont {E.}~\bibnamefont {Soergel}},\
  }\bibfield  {title} {\enquote {\bibinfo {title} {Influence of the
  inhomogeneous field at the tip on quantitative piezoresponse force
  microscopy},}\ }\href@noop {} {\bibfield  {journal} {\bibinfo  {journal}
  {Applied Physics A}\ }\textbf {\bibinfo {volume} {86}},\ \bibinfo {pages}
  {353--355} (\bibinfo {year} {2007})}\BibitemShut {NoStop}%
\bibitem [{\citenamefont {Shi}\ \emph {et~al.}(2018)\citenamefont {Shi},
  \citenamefont {Guo}, \citenamefont {Zhang},\ and\ \citenamefont
  {Guo}}]{shi-18}%
  \BibitemOpen
  \bibfield  {author} {\bibinfo {author} {\bibfnamefont {Wenhao}\ \bibnamefont
  {Shi}}, \bibinfo {author} {\bibfnamefont {Yufeng}\ \bibnamefont {Guo}},
  \bibinfo {author} {\bibfnamefont {Zhuhua}\ \bibnamefont {Zhang}}, \ and\
  \bibinfo {author} {\bibfnamefont {Wanlin}\ \bibnamefont {Guo}},\ }\bibfield
  {title} {\enquote {\bibinfo {title} {Flexoelectricity in monolayer transition
  metal dichalcogenides},}\ }\href@noop {} {\bibfield  {journal} {\bibinfo
  {journal} {The Journal of Physical Chemistry Letters}\ }\textbf {\bibinfo
  {volume} {9}},\ \bibinfo {pages} {6841--6846} (\bibinfo {year}
  {2018})}\BibitemShut {NoStop}%
\bibitem [{\citenamefont {Shi}\ \emph {et~al.}(2019)\citenamefont {Shi},
  \citenamefont {Guo}, \citenamefont {Zhang},\ and\ \citenamefont
  {Guo}}]{shi-19}%
  \BibitemOpen
  \bibfield  {author} {\bibinfo {author} {\bibfnamefont {Wenhao}\ \bibnamefont
  {Shi}}, \bibinfo {author} {\bibfnamefont {Yufeng}\ \bibnamefont {Guo}},
  \bibinfo {author} {\bibfnamefont {Zhuhua}\ \bibnamefont {Zhang}}, \ and\
  \bibinfo {author} {\bibfnamefont {Wanlin}\ \bibnamefont {Guo}},\ }\bibfield
  {title} {\enquote {\bibinfo {title} {Strain gradient mediated magnetism and
  polarization in monolayer {VSe$_2$}},}\ }\href@noop {} {\bibfield  {journal}
  {\bibinfo  {journal} {J. Phys. Chem. C}\ }\textbf {\bibinfo {volume} {123}},\
  \bibinfo {pages} {24988--24993} (\bibinfo {year} {2019})}\BibitemShut
  {NoStop}%
\bibitem [{\citenamefont {Pandey}\ \emph
  {et~al.}(2021{\natexlab{a}})\citenamefont {Pandey}, \citenamefont {Covaci},\
  and\ \citenamefont {Peeters}}]{pandey-21a}%
  \BibitemOpen
  \bibfield  {author} {\bibinfo {author} {\bibfnamefont {T.}~\bibnamefont
  {Pandey}}, \bibinfo {author} {\bibfnamefont {L.}~\bibnamefont {Covaci}}, \
  and\ \bibinfo {author} {\bibfnamefont {F.M.}\ \bibnamefont {Peeters}},\
  }\bibfield  {title} {\enquote {\bibinfo {title} {Tuning flexoelectricty and
  electronic properties of zig-zag graphene nanoribbons by
  functionalization},}\ }\href {\doibase 10.1016/j.carbon.2020.09.028}
  {\bibfield  {journal} {\bibinfo  {journal} {Carbon}\ }\textbf {\bibinfo
  {volume} {171}},\ \bibinfo {pages} {551--559} (\bibinfo {year}
  {2021}{\natexlab{a}})}\BibitemShut {NoStop}%
\bibitem [{\citenamefont {Pandey}\ \emph
  {et~al.}(2021{\natexlab{b}})\citenamefont {Pandey}, \citenamefont {Covaci},
  \citenamefont {{Milo\ifmmode \check{s}\else {\v s}\fi{}evi\ifmmode
  \acute{c}\else {\'c}\fi{}}},\ and\ \citenamefont {Peeters}}]{pandey-21b}%
  \BibitemOpen
  \bibfield  {author} {\bibinfo {author} {\bibfnamefont {T.}~\bibnamefont
  {Pandey}}, \bibinfo {author} {\bibfnamefont {L.}~\bibnamefont {Covaci}},
  \bibinfo {author} {\bibfnamefont {M.~V.}\ \bibnamefont {{Milo\ifmmode
  \check{s}\else {\v s}\fi{}evi\ifmmode \acute{c}\else {\'c}\fi{}}}}, \ and\
  \bibinfo {author} {\bibfnamefont {F.~M.}\ \bibnamefont {Peeters}},\
  }\bibfield  {title} {\enquote {\bibinfo {title} {Flexoelectricity and
  transport properties of phosphorene nanoribbons under mechanical bending},}\
  }\href {\doibase 10.1103/PhysRevB.103.235406} {\bibfield  {journal} {\bibinfo
   {journal} {Phys. Rev. B}\ }\textbf {\bibinfo {volume} {103}},\ \bibinfo
  {pages} {235406} (\bibinfo {year} {2021}{\natexlab{b}})}\BibitemShut
  {NoStop}%
\bibitem [{\citenamefont {Zhuang}\ \emph {et~al.}(2019)\citenamefont {Zhuang},
  \citenamefont {He}, \citenamefont {Javvaji},\ and\ \citenamefont
  {Park}}]{zhuang-19}%
  \BibitemOpen
  \bibfield  {author} {\bibinfo {author} {\bibfnamefont {Xiaoying}\
  \bibnamefont {Zhuang}}, \bibinfo {author} {\bibfnamefont {Bo}~\bibnamefont
  {He}}, \bibinfo {author} {\bibfnamefont {Brahmanandam}\ \bibnamefont
  {Javvaji}}, \ and\ \bibinfo {author} {\bibfnamefont {Harold~S.}\ \bibnamefont
  {Park}},\ }\bibfield  {title} {\enquote {\bibinfo {title} {Intrinsic bending
  flexoelectric constants in two-dimensional materials},}\ }\href {\doibase
  10.1103/PhysRevB.99.054105} {\bibfield  {journal} {\bibinfo  {journal} {Phys.
  Rev. B}\ }\textbf {\bibinfo {volume} {99}},\ \bibinfo {pages} {054105}
  (\bibinfo {year} {2019})}\BibitemShut {NoStop}%
\bibitem [{\citenamefont {Javvaji}\ \emph {et~al.}(2019)\citenamefont
  {Javvaji}, \citenamefont {He}, \citenamefont {Zhuang},\ and\ \citenamefont
  {Park}}]{javvaji-19}%
  \BibitemOpen
  \bibfield  {author} {\bibinfo {author} {\bibfnamefont {Brahmanandam}\
  \bibnamefont {Javvaji}}, \bibinfo {author} {\bibfnamefont {Bo}~\bibnamefont
  {He}}, \bibinfo {author} {\bibfnamefont {Xiaoying}\ \bibnamefont {Zhuang}}, \
  and\ \bibinfo {author} {\bibfnamefont {Harold~S.}\ \bibnamefont {Park}},\
  }\bibfield  {title} {\enquote {\bibinfo {title} {High flexoelectric constants
  in janus transition-metal dichalcogenides},}\ }\href {\doibase
  10.1103/PhysRevMaterials.3.125402} {\bibfield  {journal} {\bibinfo  {journal}
  {Phys. Rev. Materials}\ }\textbf {\bibinfo {volume} {3}},\ \bibinfo {pages}
  {125402} (\bibinfo {year} {2019})}\BibitemShut {NoStop}%
\bibitem [{\citenamefont {Codony}\ \emph {et~al.}(2021)\citenamefont {Codony},
  \citenamefont {Arias},\ and\ \citenamefont {Suryanarayana}}]{codony-20}%
  \BibitemOpen
  \bibfield  {author} {\bibinfo {author} {\bibfnamefont {David}\ \bibnamefont
  {Codony}}, \bibinfo {author} {\bibfnamefont {Irene}\ \bibnamefont {Arias}}, \
  and\ \bibinfo {author} {\bibfnamefont {Phanish}\ \bibnamefont
  {Suryanarayana}},\ }\bibfield  {title} {\enquote {\bibinfo {title}
  {Transversal flexoelectric coefficient for nanostructures at finite
  deformations from first principles},}\ }\href {\doibase
  10.1103/physrevmaterials.5.l030801} {\bibfield  {journal} {\bibinfo
  {journal} {Phys. Rev. Materials}\ }\textbf {\bibinfo {volume} {5}} (\bibinfo
  {year} {2021}),\ 10.1103/physrevmaterials.5.l030801}\BibitemShut {NoStop}%
\bibitem [{\citenamefont {Stengel}(2013{\natexlab{a}})}]{stengel-13b}%
  \BibitemOpen
  \bibfield  {author} {\bibinfo {author} {\bibfnamefont {M.}~\bibnamefont
  {Stengel}},\ }\bibfield  {title} {\enquote {\bibinfo {title} {Microscopic
  response to inhomogeneous deformations in curvilinear coordinates},}\
  }\href@noop {} {\bibfield  {journal} {\bibinfo  {journal} {Nature
  Communications}\ }\textbf {\bibinfo {volume} {4}},\ \bibinfo {pages} {2693}
  (\bibinfo {year} {2013}{\natexlab{a}})}\BibitemShut {NoStop}%
\bibitem [{\citenamefont {Cross}(2006)}]{cross-06}%
  \BibitemOpen
  \bibfield  {author} {\bibinfo {author} {\bibfnamefont {L.~Eric}\ \bibnamefont
  {Cross}},\ }\bibfield  {title} {\enquote {\bibinfo {title} {Flexoelectric
  effects: Charge separation in insulating solids subjected to elastic strain
  gradients},}\ }\href@noop {} {\bibfield  {journal} {\bibinfo  {journal}
  {Journal of Materials Science}\ }\textbf {\bibinfo {volume} {41}},\ \bibinfo
  {pages} {53--63} (\bibinfo {year} {2006})}\BibitemShut {NoStop}%
\bibitem [{\citenamefont {Yudin}\ and\ \citenamefont
  {Tagantsev}(2013)}]{yudin-13}%
  \BibitemOpen
  \bibfield  {author} {\bibinfo {author} {\bibfnamefont {P.~V.}\ \bibnamefont
  {Yudin}}\ and\ \bibinfo {author} {\bibfnamefont {A.~K.}\ \bibnamefont
  {Tagantsev}},\ }\bibfield  {title} {\enquote {\bibinfo {title} {Fundamentals
  of flexoelectricity in solids},}\ }\href@noop {} {\bibfield  {journal}
  {\bibinfo  {journal} {Nanotechnology}\ }\textbf {\bibinfo {volume} {24}},\
  \bibinfo {pages} {432001} (\bibinfo {year} {2013})}\BibitemShut {NoStop}%
\bibitem [{\citenamefont {Resta}(2010)}]{resta-10}%
  \BibitemOpen
  \bibfield  {author} {\bibinfo {author} {\bibfnamefont {Raffaele}\
  \bibnamefont {Resta}},\ }\bibfield  {title} {\enquote {\bibinfo {title}
  {Towards a bulk theory of flexoelectricity},}\ }\href {\doibase
  10.1103/PhysRevLett.105.127601} {\bibfield  {journal} {\bibinfo  {journal}
  {Phys. Rev. Lett.}\ }\textbf {\bibinfo {volume} {105}},\ \bibinfo {pages}
  {127601} (\bibinfo {year} {2010})}\BibitemShut {NoStop}%
\bibitem [{\citenamefont {Stengel}(2013{\natexlab{b}})}]{stengel-13a}%
  \BibitemOpen
  \bibfield  {author} {\bibinfo {author} {\bibfnamefont {M.}~\bibnamefont
  {Stengel}},\ }\bibfield  {title} {\enquote {\bibinfo {title}
  {Flexoelectricity from density-functional perturbation theory},}\ }\href@noop
  {} {\bibfield  {journal} {\bibinfo  {journal} {Phys. Rev. B}\ }\textbf
  {\bibinfo {volume} {88}},\ \bibinfo {pages} {174106} (\bibinfo {year}
  {2013}{\natexlab{b}})}\BibitemShut {NoStop}%
\bibitem [{\citenamefont {Stengel}\ and\ \citenamefont
  {Vanderbilt}(2016)}]{chapter-15}%
  \BibitemOpen
  \bibfield  {author} {\bibinfo {author} {\bibfnamefont {Massimiliano}\
  \bibnamefont {Stengel}}\ and\ \bibinfo {author} {\bibfnamefont {David}\
  \bibnamefont {Vanderbilt}},\ }\bibfield  {title} {\enquote {\bibinfo {title}
  {First-principles theory of flexoelectricity},}\ }in\ \href@noop {} {\emph
  {\bibinfo {booktitle} {Flexoelectricity in Solids From Theory to
  Applications}}},\ \bibinfo {editor} {edited by\ \bibinfo {editor}
  {\bibfnamefont {Alexander~K.}\ \bibnamefont {Tagantsev}}\ and\ \bibinfo
  {editor} {\bibfnamefont {Petr~V.}\ \bibnamefont {Yudin}}}\ (\bibinfo
  {publisher} {World Scientific Publishing Co.},\ \bibinfo {address}
  {Singapore},\ \bibinfo {year} {2016})\ Chap.~\bibinfo {chapter} {2}, pp.\
  \bibinfo {pages} {31--110}\BibitemShut {NoStop}%
\bibitem [{\citenamefont {Dreyer}\ \emph {et~al.}(2018)\citenamefont {Dreyer},
  \citenamefont {Stengel},\ and\ \citenamefont {Vanderbilt}}]{dreyer-18}%
  \BibitemOpen
  \bibfield  {author} {\bibinfo {author} {\bibfnamefont {Cyrus~E.}\
  \bibnamefont {Dreyer}}, \bibinfo {author} {\bibfnamefont {Massimiliano}\
  \bibnamefont {Stengel}}, \ and\ \bibinfo {author} {\bibfnamefont {David}\
  \bibnamefont {Vanderbilt}},\ }\bibfield  {title} {\enquote {\bibinfo {title}
  {Current-density implementation for calculating flexoelectric
  coefficients},}\ }\href {\doibase 10.1103/PhysRevB.98.075153} {\bibfield
  {journal} {\bibinfo  {journal} {Phys. Rev. B}\ }\textbf {\bibinfo {volume}
  {98}},\ \bibinfo {pages} {075153} (\bibinfo {year} {2018})}\BibitemShut
  {NoStop}%
\bibitem [{\citenamefont {Schiaffino}\ \emph {et~al.}(2019)\citenamefont
  {Schiaffino}, \citenamefont {Dreyer}, \citenamefont {Vanderbilt},\ and\
  \citenamefont {Stengel}}]{schiaffino-19}%
  \BibitemOpen
  \bibfield  {author} {\bibinfo {author} {\bibfnamefont {Andrea}\ \bibnamefont
  {Schiaffino}}, \bibinfo {author} {\bibfnamefont {Cyrus~E.}\ \bibnamefont
  {Dreyer}}, \bibinfo {author} {\bibfnamefont {David}\ \bibnamefont
  {Vanderbilt}}, \ and\ \bibinfo {author} {\bibfnamefont {Massimiliano}\
  \bibnamefont {Stengel}},\ }\bibfield  {title} {\enquote {\bibinfo {title}
  {Metric wave approach to flexoelectricity within density functional
  perturbation theory},}\ }\href {\doibase 10.1103/PhysRevB.99.085107}
  {\bibfield  {journal} {\bibinfo  {journal} {Phys. Rev. B}\ }\textbf {\bibinfo
  {volume} {99}},\ \bibinfo {pages} {085107} (\bibinfo {year}
  {2019})}\BibitemShut {NoStop}%
\bibitem [{\citenamefont {Royo}\ and\ \citenamefont {Stengel}(2019)}]{royo-19}%
  \BibitemOpen
  \bibfield  {author} {\bibinfo {author} {\bibfnamefont {Miquel}\ \bibnamefont
  {Royo}}\ and\ \bibinfo {author} {\bibfnamefont {Massimiliano}\ \bibnamefont
  {Stengel}},\ }\bibfield  {title} {\enquote {\bibinfo {title}
  {First-principles theory of spatial dispersion: Dynamical quadrupoles and
  flexoelectricity},}\ }\href {\doibase 10.1103/PhysRevX.9.021050} {\bibfield
  {journal} {\bibinfo  {journal} {Phys. Rev. X}\ }\textbf {\bibinfo {volume}
  {9}},\ \bibinfo {pages} {021050} (\bibinfo {year} {2019})}\BibitemShut
  {NoStop}%
\bibitem [{\citenamefont {Romero}\ \emph {et~al.}(2020)\citenamefont {Romero},
  \citenamefont {Allan}, \citenamefont {Amadon}, \citenamefont {Antonius},
  \citenamefont {Applencourt}, \citenamefont {Baguet}, \citenamefont {Bieder},
  \citenamefont {Bottin}, \citenamefont {Bouchet}, \citenamefont {Bousquet},
  \citenamefont {Bruneval}, \citenamefont {Brunin}, \citenamefont {Caliste},
  \citenamefont {C{\^o}t{\'e}}, \citenamefont {Denier}, \citenamefont {Dreyer},
  \citenamefont {Ghosez}, \citenamefont {Giantomassi}, \citenamefont {Gillet},
  \citenamefont {Gingras}, \citenamefont {Hamann}, \citenamefont {Hautier},
  \citenamefont {Jollet}, \citenamefont {Jomard}, \citenamefont {Martin},
  \citenamefont {Miranda}, \citenamefont {Naccarato}, \citenamefont {Petretto},
  \citenamefont {Pike}, \citenamefont {Planes}, \citenamefont {Prokhorenko},
  \citenamefont {Rangel}, \citenamefont {Ricci}, \citenamefont {Rignanese},
  \citenamefont {Royo}, \citenamefont {Stengel}, \citenamefont {Torrent},
  \citenamefont {van Setten}, \citenamefont {{Van Troeye}}, \citenamefont
  {Verstraete}, \citenamefont {Wiktor}, \citenamefont {Zwanziger},\ and\
  \citenamefont {Gonze}}]{romero-20}%
  \BibitemOpen
  \bibfield  {author} {\bibinfo {author} {\bibfnamefont {Aldo~H.}\ \bibnamefont
  {Romero}}, \bibinfo {author} {\bibfnamefont {Douglas~C.}\ \bibnamefont
  {Allan}}, \bibinfo {author} {\bibfnamefont {Bernard}\ \bibnamefont {Amadon}},
  \bibinfo {author} {\bibfnamefont {Gabriel}\ \bibnamefont {Antonius}},
  \bibinfo {author} {\bibfnamefont {Thomas}\ \bibnamefont {Applencourt}},
  \bibinfo {author} {\bibfnamefont {Lucas}\ \bibnamefont {Baguet}}, \bibinfo
  {author} {\bibfnamefont {Jordan}\ \bibnamefont {Bieder}}, \bibinfo {author}
  {\bibfnamefont {Fran{\c c}ois}\ \bibnamefont {Bottin}}, \bibinfo {author}
  {\bibfnamefont {Johann}\ \bibnamefont {Bouchet}}, \bibinfo {author}
  {\bibfnamefont {Eric}\ \bibnamefont {Bousquet}}, \bibinfo {author}
  {\bibfnamefont {Fabien}\ \bibnamefont {Bruneval}}, \bibinfo {author}
  {\bibfnamefont {Guillaume}\ \bibnamefont {Brunin}}, \bibinfo {author}
  {\bibfnamefont {Damien}\ \bibnamefont {Caliste}}, \bibinfo {author}
  {\bibfnamefont {Michel}\ \bibnamefont {C{\^o}t{\'e}}}, \bibinfo {author}
  {\bibfnamefont {Jules}\ \bibnamefont {Denier}}, \bibinfo {author}
  {\bibfnamefont {Cyrus}\ \bibnamefont {Dreyer}}, \bibinfo {author}
  {\bibfnamefont {Philippe}\ \bibnamefont {Ghosez}}, \bibinfo {author}
  {\bibfnamefont {Matteo}\ \bibnamefont {Giantomassi}}, \bibinfo {author}
  {\bibfnamefont {Yannick}\ \bibnamefont {Gillet}}, \bibinfo {author}
  {\bibfnamefont {Olivier}\ \bibnamefont {Gingras}}, \bibinfo {author}
  {\bibfnamefont {Donald~R.}\ \bibnamefont {Hamann}}, \bibinfo {author}
  {\bibfnamefont {Geoffroy}\ \bibnamefont {Hautier}}, \bibinfo {author}
  {\bibfnamefont {Fran{\c c}ois}\ \bibnamefont {Jollet}}, \bibinfo {author}
  {\bibfnamefont {G{\'e}rald}\ \bibnamefont {Jomard}}, \bibinfo {author}
  {\bibfnamefont {Alexandre}\ \bibnamefont {Martin}}, \bibinfo {author}
  {\bibfnamefont {Henrique P.~C.}\ \bibnamefont {Miranda}}, \bibinfo {author}
  {\bibfnamefont {Francesco}\ \bibnamefont {Naccarato}}, \bibinfo {author}
  {\bibfnamefont {Guido}\ \bibnamefont {Petretto}}, \bibinfo {author}
  {\bibfnamefont {Nicholas~A.}\ \bibnamefont {Pike}}, \bibinfo {author}
  {\bibfnamefont {Valentin}\ \bibnamefont {Planes}}, \bibinfo {author}
  {\bibfnamefont {Sergei}\ \bibnamefont {Prokhorenko}}, \bibinfo {author}
  {\bibfnamefont {Tonatiuh}\ \bibnamefont {Rangel}}, \bibinfo {author}
  {\bibfnamefont {Fabio}\ \bibnamefont {Ricci}}, \bibinfo {author}
  {\bibfnamefont {Gian-Marco}\ \bibnamefont {Rignanese}}, \bibinfo {author}
  {\bibfnamefont {Miquel}\ \bibnamefont {Royo}}, \bibinfo {author}
  {\bibfnamefont {Massimiliano}\ \bibnamefont {Stengel}}, \bibinfo {author}
  {\bibfnamefont {Marc}\ \bibnamefont {Torrent}}, \bibinfo {author}
  {\bibfnamefont {Michiel~J.}\ \bibnamefont {van Setten}}, \bibinfo {author}
  {\bibfnamefont {Benoit}\ \bibnamefont {{Van Troeye}}}, \bibinfo {author}
  {\bibfnamefont {Matthieu~J.}\ \bibnamefont {Verstraete}}, \bibinfo {author}
  {\bibfnamefont {Julia}\ \bibnamefont {Wiktor}}, \bibinfo {author}
  {\bibfnamefont {Josef~W.}\ \bibnamefont {Zwanziger}}, \ and\ \bibinfo
  {author} {\bibfnamefont {Xavier}\ \bibnamefont {Gonze}},\ }\bibfield  {title}
  {\enquote {\bibinfo {title} {Abinit: Overview and focus on selected
  capabilities},}\ }\href {\doibase 10.1063/1.5144261} {\bibfield  {journal}
  {\bibinfo  {journal} {The Journal of Chemical Physics}\ }\textbf {\bibinfo
  {volume} {152}},\ \bibinfo {pages} {124102} (\bibinfo {year}
  {2020})}\BibitemShut {NoStop}%
\bibitem [{\citenamefont {Martin}(1972)}]{martin-72}%
  \BibitemOpen
  \bibfield  {author} {\bibinfo {author} {\bibfnamefont {Richard~M.}\
  \bibnamefont {Martin}},\ }\bibfield  {title} {\enquote {\bibinfo {title}
  {Piezoelectricity},}\ }\href {\doibase 10.1103/PhysRevB.5.1607} {\bibfield
  {journal} {\bibinfo  {journal} {Phys. Rev. B}\ }\textbf {\bibinfo {volume}
  {5}},\ \bibinfo {pages} {1607--1613} (\bibinfo {year} {1972})}\BibitemShut
  {NoStop}%
\bibitem [{\citenamefont {{See Supplemental Material at http://link for more
  information about theoretical derivations, computational details and results,
  which includes Refs. [39-48].}}()}]{supplemental_prl}%
  \BibitemOpen
  \bibfield  {author} {\bibinfo {author} {\bibnamefont {{See Supplemental
  Material at http://link for more information about theoretical derivations,
  computational details and results, which includes Refs. [39-48].}}},\
  }\href@noop {} {}\BibitemShut {NoStop}%
\bibitem [{\citenamefont {Stengel}(2014)}]{stengel-14}%
  \BibitemOpen
  \bibfield  {author} {\bibinfo {author} {\bibfnamefont {Massimiliano}\
  \bibnamefont {Stengel}},\ }\bibfield  {title} {\enquote {\bibinfo {title}
  {Surface control of flexoelectricity},}\ }\href {\doibase
  10.1103/PhysRevB.90.201112} {\bibfield  {journal} {\bibinfo  {journal} {Phys.
  Rev. B}\ }\textbf {\bibinfo {volume} {90}},\ \bibinfo {pages} {201112}
  (\bibinfo {year} {2014})}\BibitemShut {NoStop}%
\bibitem [{\citenamefont {Hong}\ and\ \citenamefont
  {Vanderbilt}(2013)}]{hong-13}%
  \BibitemOpen
  \bibfield  {author} {\bibinfo {author} {\bibfnamefont {Jiawang}\ \bibnamefont
  {Hong}}\ and\ \bibinfo {author} {\bibfnamefont {David}\ \bibnamefont
  {Vanderbilt}},\ }\bibfield  {title} {\enquote {\bibinfo {title}
  {First-principles theory and calculation of flexoelectricity},}\ }\href
  {\doibase 10.1103/PhysRevB.88.174107} {\bibfield  {journal} {\bibinfo
  {journal} {Phys. Rev. B}\ }\textbf {\bibinfo {volume} {88}},\ \bibinfo
  {pages} {174107} (\bibinfo {year} {2013})}\BibitemShut {NoStop}%
\bibitem [{\citenamefont {Baroni}\ \emph {et~al.}(2001)\citenamefont {Baroni},
  \citenamefont {de~Gironcoli},\ and\ \citenamefont {Corso}}]{baroni-01}%
  \BibitemOpen
  \bibfield  {author} {\bibinfo {author} {\bibfnamefont {S.}~\bibnamefont
  {Baroni}}, \bibinfo {author} {\bibfnamefont {S.}~\bibnamefont
  {de~Gironcoli}}, \ and\ \bibinfo {author} {\bibfnamefont {A.~Dal}\
  \bibnamefont {Corso}},\ }\bibfield  {title} {\enquote {\bibinfo {title}
  {Phonons and related crystal properties from density-functional perturbation
  theory},}\ }\href@noop {} {\bibfield  {journal} {\bibinfo  {journal} {Rev.
  Mod. Phys.}\ }\textbf {\bibinfo {volume} {73}},\ \bibinfo {pages} {515}
  (\bibinfo {year} {2001})}\BibitemShut {NoStop}%
\bibitem [{\citenamefont {Gonze}\ and\ \citenamefont {Lee}(1997)}]{gonze-97b}%
  \BibitemOpen
  \bibfield  {author} {\bibinfo {author} {\bibfnamefont {X.}~\bibnamefont
  {Gonze}}\ and\ \bibinfo {author} {\bibfnamefont {C.}~\bibnamefont {Lee}},\
  }\bibfield  {title} {\enquote {\bibinfo {title} {Dynamical matrices, {Born}
  effective charges, dielectric permittivity tensors, and interatomic force
  constants from density-functional perturbation theory},}\ }\href@noop {}
  {\bibfield  {journal} {\bibinfo  {journal} {Phys. Rev. B}\ }\textbf {\bibinfo
  {volume} {55}},\ \bibinfo {pages} {10355} (\bibinfo {year}
  {1997})}\BibitemShut {NoStop}%
\bibitem [{\citenamefont {Gonze}\ \emph {et~al.}(2009)\citenamefont {Gonze},
  \citenamefont {Amadon}, \citenamefont {Anglade}, \citenamefont {Beuken},
  \citenamefont {Bottin}, \citenamefont {Boulanger}, \citenamefont {Bruneval},
  \citenamefont {Caliste}, \citenamefont {Caracas}, \citenamefont
  {{C\^ot{\'e}}}, \citenamefont {Deutsch}, \citenamefont {Genovese},
  \citenamefont {Ghosez}, \citenamefont {Giantomassi}, \citenamefont
  {Goedecker}, \citenamefont {Hamann}, \citenamefont {Hermet}, \citenamefont
  {Jollet}, \citenamefont {Jomard}, \citenamefont {Leroux}, \citenamefont
  {Mancini}, \citenamefont {Mazevet}, \citenamefont {Oliveira}, \citenamefont
  {Onida}, \citenamefont {Pouillon}, \citenamefont {Rangel}, \citenamefont
  {Rignanese}, \citenamefont {Sangalli}, \citenamefont {Shaltaf}, \citenamefont
  {Torrent}, \citenamefont {Verstraete}, \citenamefont {Zerah},\ and\
  \citenamefont {Zwanziger}}]{abinit}%
  \BibitemOpen
  \bibfield  {author} {\bibinfo {author} {\bibfnamefont {X.}~\bibnamefont
  {Gonze}}, \bibinfo {author} {\bibfnamefont {B.}~\bibnamefont {Amadon}},
  \bibinfo {author} {\bibfnamefont {P.-M.}\ \bibnamefont {Anglade}}, \bibinfo
  {author} {\bibfnamefont {J.-M.}\ \bibnamefont {Beuken}}, \bibinfo {author}
  {\bibfnamefont {F.}~\bibnamefont {Bottin}}, \bibinfo {author} {\bibfnamefont
  {P.}~\bibnamefont {Boulanger}}, \bibinfo {author} {\bibfnamefont
  {F.}~\bibnamefont {Bruneval}}, \bibinfo {author} {\bibfnamefont
  {D.}~\bibnamefont {Caliste}}, \bibinfo {author} {\bibfnamefont
  {R.}~\bibnamefont {Caracas}}, \bibinfo {author} {\bibfnamefont
  {M.}~\bibnamefont {{C\^ot{\'e}}}}, \bibinfo {author} {\bibfnamefont
  {T.}~\bibnamefont {Deutsch}}, \bibinfo {author} {\bibfnamefont
  {L.}~\bibnamefont {Genovese}}, \bibinfo {author} {\bibfnamefont {Ph.}\
  \bibnamefont {Ghosez}}, \bibinfo {author} {\bibfnamefont {M.}~\bibnamefont
  {Giantomassi}}, \bibinfo {author} {\bibfnamefont {S.}~\bibnamefont
  {Goedecker}}, \bibinfo {author} {\bibfnamefont {D.R.}\ \bibnamefont
  {Hamann}}, \bibinfo {author} {\bibfnamefont {P.}~\bibnamefont {Hermet}},
  \bibinfo {author} {\bibfnamefont {F.}~\bibnamefont {Jollet}}, \bibinfo
  {author} {\bibfnamefont {G.}~\bibnamefont {Jomard}}, \bibinfo {author}
  {\bibfnamefont {S.}~\bibnamefont {Leroux}}, \bibinfo {author} {\bibfnamefont
  {M.}~\bibnamefont {Mancini}}, \bibinfo {author} {\bibfnamefont
  {S.}~\bibnamefont {Mazevet}}, \bibinfo {author} {\bibfnamefont {M.J.T.}\
  \bibnamefont {Oliveira}}, \bibinfo {author} {\bibfnamefont {G.}~\bibnamefont
  {Onida}}, \bibinfo {author} {\bibfnamefont {Y.}~\bibnamefont {Pouillon}},
  \bibinfo {author} {\bibfnamefont {T.}~\bibnamefont {Rangel}}, \bibinfo
  {author} {\bibfnamefont {G.-M.}\ \bibnamefont {Rignanese}}, \bibinfo {author}
  {\bibfnamefont {D.}~\bibnamefont {Sangalli}}, \bibinfo {author}
  {\bibfnamefont {R.}~\bibnamefont {Shaltaf}}, \bibinfo {author} {\bibfnamefont
  {M.}~\bibnamefont {Torrent}}, \bibinfo {author} {\bibfnamefont {M.J.}\
  \bibnamefont {Verstraete}}, \bibinfo {author} {\bibfnamefont
  {G.}~\bibnamefont {Zerah}}, \ and\ \bibinfo {author} {\bibfnamefont {J.W.}\
  \bibnamefont {Zwanziger}},\ }\bibfield  {title} {\enquote {\bibinfo {title}
  {{ABINIT}: {F}irst-principles approach to material and nanosystem
  properties},}\ }\href@noop {} {\bibfield  {journal} {\bibinfo  {journal}
  {Computer Phys. Commun.}\ }\textbf {\bibinfo {volume} {180}},\ \bibinfo
  {pages} {2582--2615} (\bibinfo {year} {2009})}\BibitemShut {NoStop}%
\bibitem [{\citenamefont {Amorim}\ and\ \citenamefont
  {Guinea}(2013)}]{amorim-13}%
  \BibitemOpen
  \bibfield  {author} {\bibinfo {author} {\bibfnamefont {Bruno}\ \bibnamefont
  {Amorim}}\ and\ \bibinfo {author} {\bibfnamefont {Francisco}\ \bibnamefont
  {Guinea}},\ }\bibfield  {title} {\enquote {\bibinfo {title} {Flexural mode of
  graphene on a substrate},}\ }\href {\doibase 10.1103/PhysRevB.88.115418}
  {\bibfield  {journal} {\bibinfo  {journal} {Phys. Rev. B}\ }\textbf {\bibinfo
  {volume} {88}},\ \bibinfo {pages} {115418} (\bibinfo {year}
  {2013})}\BibitemShut {NoStop}%
\bibitem [{\citenamefont {Chu}\ \emph {et~al.}(2009)\citenamefont {Chu},
  \citenamefont {Zhu}, \citenamefont {Li},\ and\ \citenamefont
  {Cross}}]{chu-09}%
  \BibitemOpen
  \bibfield  {author} {\bibinfo {author} {\bibfnamefont {Baojin}\ \bibnamefont
  {Chu}}, \bibinfo {author} {\bibfnamefont {Wenyi}\ \bibnamefont {Zhu}},
  \bibinfo {author} {\bibfnamefont {Nan}\ \bibnamefont {Li}}, \ and\ \bibinfo
  {author} {\bibfnamefont {L.~Eric}\ \bibnamefont {Cross}},\ }\bibfield
  {title} {\enquote {\bibinfo {title} {Flexure mode flexoelectric piezoelectric
  composites},}\ }\href {\doibase 10.1063/1.3262495} {\bibfield  {journal}
  {\bibinfo  {journal} {Journal of Applied Physics}\ }\textbf {\bibinfo
  {volume} {106}},\ \bibinfo {pages} {104109} (\bibinfo {year}
  {2009})}\BibitemShut {NoStop}%
\bibitem [{\citenamefont {Royo}\ \emph {et~al.}(2020)\citenamefont {Royo},
  \citenamefont {Hahn},\ and\ \citenamefont {Stengel}}]{royo-20}%
  \BibitemOpen
  \bibfield  {author} {\bibinfo {author} {\bibfnamefont {Miquel}\ \bibnamefont
  {Royo}}, \bibinfo {author} {\bibfnamefont {Konstanze~R}\ \bibnamefont
  {Hahn}}, \ and\ \bibinfo {author} {\bibfnamefont {Massimiliano}\ \bibnamefont
  {Stengel}},\ }\bibfield  {title} {\enquote {\bibinfo {title} {Using high
  multipolar orders to reconstruct the sound velocity in piezoelectrics from
  lattice dynamics},}\ }\href@noop {} {\bibfield  {journal} {\bibinfo
  {journal} {Phys. Rev. Lett.}\ }\textbf {\bibinfo {volume} {125}},\ \bibinfo
  {pages} {217602} (\bibinfo {year} {2020})}\BibitemShut {NoStop}%
\bibitem [{\citenamefont {Hamann}(2013)}]{HamannPRB13}%
  \BibitemOpen
  \bibfield  {author} {\bibinfo {author} {\bibfnamefont {D.~R.}\ \bibnamefont
  {Hamann}},\ }\bibfield  {title} {\enquote {\bibinfo {title} {Optimized
  norm-conserving vanderbilt pseudopotentials},}\ }\href {\doibase
  10.1103/PhysRevB.88.085117} {\bibfield  {journal} {\bibinfo  {journal} {Phys.
  Rev. B}\ }\textbf {\bibinfo {volume} {88}},\ \bibinfo {pages} {085117}
  (\bibinfo {year} {2013})}\BibitemShut {NoStop}%
\bibitem [{\citenamefont {{van Setten}}\ \emph {et~al.}(2018)\citenamefont
  {{van Setten}}, \citenamefont {Giantomassi}, \citenamefont {Bousquet},
  \citenamefont {Verstraete}, \citenamefont {Hamann}, \citenamefont {Gonze},\
  and\ \citenamefont {Rignanese}}]{pseudodojo}%
  \BibitemOpen
  \bibfield  {author} {\bibinfo {author} {\bibfnamefont {M.J.}\ \bibnamefont
  {{van Setten}}}, \bibinfo {author} {\bibfnamefont {M.}~\bibnamefont
  {Giantomassi}}, \bibinfo {author} {\bibfnamefont {E.}~\bibnamefont
  {Bousquet}}, \bibinfo {author} {\bibfnamefont {M.J.}\ \bibnamefont
  {Verstraete}}, \bibinfo {author} {\bibfnamefont {D.R.}\ \bibnamefont
  {Hamann}}, \bibinfo {author} {\bibfnamefont {X.}~\bibnamefont {Gonze}}, \
  and\ \bibinfo {author} {\bibfnamefont {G.-M.}\ \bibnamefont {Rignanese}},\
  }\bibfield  {title} {\enquote {\bibinfo {title} {The {PseudoDojo}: Training
  and grading a 85 element optimized norm-conserving pseudopotential table},}\
  }\href {\doibase 10.1016/j.cpc.2018.01.012} {\bibfield  {journal} {\bibinfo
  {journal} {Comp. Phys. Comm.}\ }\textbf {\bibinfo {volume} {226}},\ \bibinfo
  {pages} {39--54} (\bibinfo {year} {2018})}\BibitemShut {NoStop}%
\bibitem [{\citenamefont {Novoselov}\ \emph {et~al.}(2016)\citenamefont
  {Novoselov}, \citenamefont {Mishchenko}, \citenamefont {Carvalho},\ and\
  \citenamefont {{Castro Neto}}}]{novoselov-16}%
  \BibitemOpen
  \bibfield  {author} {\bibinfo {author} {\bibfnamefont {K.S.}\ \bibnamefont
  {Novoselov}}, \bibinfo {author} {\bibfnamefont {A.}~\bibnamefont
  {Mishchenko}}, \bibinfo {author} {\bibfnamefont {A.}~\bibnamefont
  {Carvalho}}, \ and\ \bibinfo {author} {\bibfnamefont {A.H.}\ \bibnamefont
  {{Castro Neto}}},\ }\bibfield  {title} {\enquote {\bibinfo {title} {2d
  materials and van der waals heterostructures},}\ }\href {\doibase
  10.1126/science.aac9439} {\bibfield  {journal} {\bibinfo  {journal}
  {Science}\ }\textbf {\bibinfo {volume} {353}},\ \bibinfo {pages} {aac9439}
  (\bibinfo {year} {2016})}\BibitemShut {NoStop}%
\bibitem [{\citenamefont {Pike}\ \emph {et~al.}(2017)\citenamefont {Pike},
  \citenamefont {{Van Troeye}}, \citenamefont {Dewandre}, \citenamefont
  {Petretto}, \citenamefont {Gonze}, \citenamefont {Rignanese},\ and\
  \citenamefont {Verstraete}}]{pike-17}%
  \BibitemOpen
  \bibfield  {author} {\bibinfo {author} {\bibfnamefont {Nicholas~A.}\
  \bibnamefont {Pike}}, \bibinfo {author} {\bibfnamefont {Benoit}\ \bibnamefont
  {{Van Troeye}}}, \bibinfo {author} {\bibfnamefont {Antoine}\ \bibnamefont
  {Dewandre}}, \bibinfo {author} {\bibfnamefont {Guido}\ \bibnamefont
  {Petretto}}, \bibinfo {author} {\bibfnamefont {Xavier}\ \bibnamefont
  {Gonze}}, \bibinfo {author} {\bibfnamefont {Gian-Marco}\ \bibnamefont
  {Rignanese}}, \ and\ \bibinfo {author} {\bibfnamefont {Matthieu~J.}\
  \bibnamefont {Verstraete}},\ }\bibfield  {title} {\enquote {\bibinfo {title}
  {Origin of the counterintuitive dynamic charge in the transition metal
  dichalcogenides},}\ }\href {\doibase 10.1103/PhysRevB.95.201106} {\bibfield
  {journal} {\bibinfo  {journal} {Phys. Rev. B}\ }\textbf {\bibinfo {volume}
  {95}},\ \bibinfo {pages} {201106} (\bibinfo {year} {2017})}\BibitemShut
  {NoStop}%
\bibitem [{\citenamefont {Perdew}\ \emph {et~al.}(1996)\citenamefont {Perdew},
  \citenamefont {Burke},\ and\ \citenamefont {Ernzerhof}}]{pbe}%
  \BibitemOpen
  \bibfield  {author} {\bibinfo {author} {\bibfnamefont {John~P.}\ \bibnamefont
  {Perdew}}, \bibinfo {author} {\bibfnamefont {Kieron}\ \bibnamefont {Burke}},
  \ and\ \bibinfo {author} {\bibfnamefont {Matthias}\ \bibnamefont
  {Ernzerhof}},\ }\bibfield  {title} {\enquote {\bibinfo {title} {Generalized
  gradient approximation made simple},}\ }\href {\doibase
  10.1103/physrevlett.77.3865} {\bibfield  {journal} {\bibinfo  {journal}
  {Phys. Rev. Lett.}\ }\textbf {\bibinfo {volume} {77}},\ \bibinfo {pages}
  {3865--3868} (\bibinfo {year} {1996})}\BibitemShut {NoStop}%
\bibitem [{\citenamefont {Hamann}\ \emph {et~al.}(2005)\citenamefont {Hamann},
  \citenamefont {Rabe},\ and\ \citenamefont {Vanderbilt}}]{hamann_gga}%
  \BibitemOpen
  \bibfield  {author} {\bibinfo {author} {\bibfnamefont {D.R.}\ \bibnamefont
  {Hamann}}, \bibinfo {author} {\bibfnamefont {Karin~M.}\ \bibnamefont {Rabe}},
  \ and\ \bibinfo {author} {\bibfnamefont {David}\ \bibnamefont {Vanderbilt}},\
  }\bibfield  {title} {\enquote {\bibinfo {title}
  {Generalized-gradient-functional treatment of strain in density-functional
  perturbation theory},}\ }\href {\doibase 10.1103/physrevb.72.033102}
  {\bibfield  {journal} {\bibinfo  {journal} {Phys. Rev. B}\ }\textbf {\bibinfo
  {volume} {72}} (\bibinfo {year} {2005}),\
  10.1103/physrevb.72.033102}\BibitemShut {NoStop}%
\bibitem [{\citenamefont {Jain}\ \emph {et~al.}(2013)\citenamefont {Jain},
  \citenamefont {Ong}, \citenamefont {Hautier}, \citenamefont {Chen},
  \citenamefont {Richards}, \citenamefont {Dacek}, \citenamefont {Cholia},
  \citenamefont {Gunter}, \citenamefont {Skinner}, \citenamefont {Ceder},\ and\
  \citenamefont {Persson}}]{Jain-13}%
  \BibitemOpen
  \bibfield  {author} {\bibinfo {author} {\bibfnamefont {Anubhav}\ \bibnamefont
  {Jain}}, \bibinfo {author} {\bibfnamefont {Shyue~Ping}\ \bibnamefont {Ong}},
  \bibinfo {author} {\bibfnamefont {Geoffroy}\ \bibnamefont {Hautier}},
  \bibinfo {author} {\bibfnamefont {Wei}\ \bibnamefont {Chen}}, \bibinfo
  {author} {\bibfnamefont {William~Davidson}\ \bibnamefont {Richards}},
  \bibinfo {author} {\bibfnamefont {Stephen}\ \bibnamefont {Dacek}}, \bibinfo
  {author} {\bibfnamefont {Shreyas}\ \bibnamefont {Cholia}}, \bibinfo {author}
  {\bibfnamefont {Dan}\ \bibnamefont {Gunter}}, \bibinfo {author}
  {\bibfnamefont {David}\ \bibnamefont {Skinner}}, \bibinfo {author}
  {\bibfnamefont {Gerbrand}\ \bibnamefont {Ceder}}, \ and\ \bibinfo {author}
  {\bibfnamefont {Kristin~A.}\ \bibnamefont {Persson}},\ }\bibfield  {title}
  {\enquote {\bibinfo {title} {The materials project: A materials genome
  approach to accelerating materials innovation},}\ }\href {\doibase
  10.1063/1.4812323} {\bibfield  {journal} {\bibinfo  {journal} {APL
  Materials}\ }\textbf {\bibinfo {volume} {1}},\ \bibinfo {pages} {011002}
  (\bibinfo {year} {2013})}\BibitemShut {NoStop}%
\bibitem [{\citenamefont {Royo}\ and\ \citenamefont
  {Stengel}(2020)}]{screen_2D}%
  \BibitemOpen
  \bibfield  {author} {\bibinfo {author} {\bibfnamefont {Miquel}\ \bibnamefont
  {Royo}}\ and\ \bibinfo {author} {\bibfnamefont {Massimiliano}\ \bibnamefont
  {Stengel}},\ }\bibfield  {title} {\enquote {\bibinfo {title} {Exact
  long-range dielectric screening and interatomic force constants in quasi-2d
  crystals},}\ }\href@noop {} {\bibfield  {journal} {\bibinfo  {journal}
  {Physical Review X (accepted), arXiv preprint arXiv:2012.07961}\ } (\bibinfo
  {year} {2020})}\BibitemShut {NoStop}%
\bibitem [{\citenamefont {Kumar}\ and\ \citenamefont
  {Suryanarayana}(2020)}]{kumar-20}%
  \BibitemOpen
  \bibfield  {author} {\bibinfo {author} {\bibfnamefont {Shashikant}\
  \bibnamefont {Kumar}}\ and\ \bibinfo {author} {\bibfnamefont {Phanish}\
  \bibnamefont {Suryanarayana}},\ }\bibfield  {title} {\enquote {\bibinfo
  {title} {Bending moduli for forty-four select atomic monolayers from first
  principles},}\ }\href {\doibase 10.1088/1361-6528/aba2a2} {\bibfield
  {journal} {\bibinfo  {journal} {Nanotechnology}\ }\textbf {\bibinfo {volume}
  {31}},\ \bibinfo {pages} {43LT01} (\bibinfo {year} {2020})}\BibitemShut
  {NoStop}%
\end{thebibliography}%

\end{document}